\documentclass[preprint,review,12pt]{elsarticle}

\usepackage{amssymb}

\usepackage{hyperref}
\usepackage{amsmath}
\usepackage{cleveref}
\usepackage{booktabs}
\usepackage{caption,setspace}
\usepackage{tabularx}
\usepackage[export]{adjustbox}
\usepackage{tikz}
\usepackage{geometry}
\usepackage{wrapfig} 
\usepackage{flushend} 
\usepackage{algorithm}
\usepackage{algpseudocode}
\usepackage{xcolor}
\usepackage{multirow}

\usepackage{helvet}  

\hypersetup{pdfauthor={Name}}

\makeatletter
\newcommand{\ssymbol}[1]{^{\@fnsymbol{#1}}}
\makeatother

\usepackage{array}
\newcolumntype{P}[1]{>{\centering\arraybackslash}p{#1}}

\AtBeginDocument{
    \crefformat{equation}{#2Eq.~#1#3}
    \crefformat{figure}{#2Fig.~#1#3}
    \crefformat{table}{#2Tab.~#1#3}
}

\journal{arXiv} 

\begin{document}
\begin{sloppypar}

\begin{frontmatter}

\title{Segmenting Medical Images with Limited Data}


\author[a]{Zhaoshan Liu\corref{cor1}}
\ead{e0575844@u.nus.edu}

\author[a,b]{Qiujie Lv\corref{cor1}}
\ead{lvqj5@mail2.sysu.edu.cn} 

\author[c]{Chau Hung Lee}
\ead{chau_hung_lee@ttsh.com.sg}

\author[a]{Lei Shen\corref{cor2}}
\ead{mpeshel@nus.edu.sg}

\cortext[cor1]{Equal contribution}

\cortext[cor2]{Corresponding author}

\address[a]{Department of Mechanical Engineering, National University of Singapore, 9 Engineering Drive 1, Singapore, 117575, Singapore}

\address[b]{School of Intelligent Systems Engineering, Sun Yat-sen University, No.66, Gongchang Road, Guangming District, 518107, China}

\address[c]{Department of Radiology, Tan Tock Seng Hospital, 11 Jalan Tan Tock Seng, Singapore, 308433, Singapore}


\begin{abstract}
While computer vision has proven valuable for medical image segmentation, its application faces challenges such as limited dataset sizes and the complexity of effectively leveraging unlabeled images. To address these challenges, we present a novel semi-supervised, consistency-based approach termed the data-efficient medical segmenter (DEMS). The DEMS features an encoder-decoder architecture and incorporates the developed online automatic augmenter (OAA) and residual robustness enhancement (RRE) blocks. The OAA augments input data with various image transformations, thereby diversifying the dataset to improve the generalization ability. The RRE enriches feature diversity and introduces perturbations to create varied inputs for different decoders, thereby providing enhanced variability. Moreover, we introduce a sensitive loss to further enhance consistency across different decoders and stabilize the training process. Extensive experimental results on both our own and three public datasets affirm the effectiveness of DEMS. Under extreme data shortage scenarios, our DEMS achieves 16.85\% and 10.37\% improvement in dice score compared with the U-Net and top-performed state-of-the-art method, respectively. Given its superior data efficiency, DEMS could present significant advancements in medical segmentation under small data regimes. The project homepage can be accessed at \href{https://github.com/NUS-Tim/DEMS}{https://github.com/NUS-Tim/DEMS}.
\end{abstract}



\end{frontmatter}



\section{Introduction}
\label{1}

Recently, there has been a rapid advancement in artificial intelligence \citep{hu2023detdo, zare2023global, abualigah2023modified}, with computer vision emerging as a particularly prominent area of research. It has witnessed remarkable progress in a variety of applications \citep{chen2024dual, yelleni2024monte, liu2023recent, guo2023survival}, including facial recognition, autonomous navigation, and medical image segmentation. Medical image segmentation leverages the neural network model to generate high-accuracy mask prediction and has achieved significant success in various scenarios \citep{zhou2023volumetric, xie2022context, guo2023causal}. Although model assistance greatly enhances medical diagnosis, its usage faces unique challenges. In particular, training the neural network often requires large labeled datasets, yet collecting medical images is not as straightforward as collecting natural images \citep{krizhevsky2017imagenet}. This can be attributed to several factors, including cost and privacy concerns \citep{liu2023gsda}. First, acquiring medical images is time-consuming and requires specialized equipment and expert intervention. Additionally, in some scenarios, there may not be a sufficient number of patients available for data collection. Second, scanned images require additional labeling, which further increases the time cost. Finally, patient privacy \citep{murdoch2021privacy} must be considered, making a significant portion of datasets publicly unavailable. In this context, the number of accessible images may be limited, and a substantial amount of unlabeled images may remain underutilized for model training.

To facilitate effective model training with limited data and bolster generalizability, a multitude of data augmentation (DA) methods have been presented. Prevalent DA methods can be categorized into three types, including conventional DA, generative adversarial network (GAN)-based DA, and automatic DA. Conventional DA leverages various image transformations such as flipping, rotation, contrast adjustment, and scaling \citep{eisenmann2023winner} to generate augmented variants. This approach is the most commonly implemented \citep{li2023hal, zhao2023mskd, isensee2021nnu}, yet the augmentation pipeline design heavily relies on experience. The GAN-based DA leverages GAN \citep{goodfellow2014generative} to synthesize artificial images and enlarge datasets \citep{chai2022synthetic, li2021semantic, kugelman2020dual}. This method, while more versatile, presents challenges such as variable synthetic quality, extensive training times \citep{feng2021gans, zhou2021jittor}, and strong dependence of model performance on dataset size \citep{feng2021gans}. Automatic DA \citep{cubuk2019autoaugment} integrates multiple conventional DA transformations to enhance data diversity \citep{yang2019searching, lyu2022aadg, qin2020automatic}. However, its application faces various challenges, including extensive computational demands, complex stage design, and potential model confounding \citep{yun2019cutmix}.

Semi-supervised learning (SSL) has been proposed to effectively utilize unlabeled images \citep{jiao2022learning} and has garnered significant attention in the field of medical image segmentation. Its applications predominantly rely on various intrinsic principles such as consistency regularization \citep{ouali2020semi}, pseudo-labeling \citep{li2023semi}, and prior knowledge \citep{zheng2019semi}, with consistency regularization being the most prevalent. The consistency-based method encourages the model to generate consistent outputs when taking inputs that undergo various operations or perturbations, thereby enhancing its generalization ability. A significant number of consistency-based methods adopt the mean teacher (MT) architecture \citep{tarvainen2017mean} as their foundation \citep{zhang2023uncertainty, lei2022semi, xu2021shadow}. The MT-based methods comprise a teacher subnet and a student subnet, where the parameters of the teacher network are updated from the student network using the exponential moving average algorithm. However, challenges arise with the MT-based approaches, such as the constraint on teacher network performance and limited model variability \citep{wang2023mcf}. In addition to the MT-based architecture, various innovative methods demonstrate remarkable performance, such as those based on encoder-decoder architecture \citep{tang2023multi, wu2021semi, wu2022mutual} and GAN architecture \citep{zhai2022ass}. With sophisticated network and loss design, these approaches have demonstrated outstanding performance, while challenges such as variability restriction and stability limitation \citep{xu2022pca} persist.

To bolster model performance with limited data comprising unlabeled images, we propose a data-efficient medical segmenter (DEMS). The DEMS employs an encoder-decoder architecture and comprises our developed online automatic augmenter (OAA) and residual robustness enhancement (RRE) blocks. The OAA diversifies visual input with varying transformations, thereby providing enhanced data diversity to bolster generalization ability. The RRE block enriches feature diversity and introduces perturbations to generate varied inputs for different decoders and thus offers greater variability. Additionally, we introduce a novel sensitivity loss to enhance consistency across different decoders and stabilize model training. We perform extensive experiments on our own and three public datasets and the results show that DEMS outperforms state-of-the-art (SOTA) methods. Furthermore, DEMS exhibits greater performance leadership under severe data regimes, thereby highlighting its exceptional data efficiency. To sum up, our main contributions are:
\begin{itemize}
    \item We propose a data-efficient medical segmenter (DEMS), a novel semi-supervised segmentation approach that effectively utilizes limited and unlabeled data and features superior performance in small data scenarios.
    \item We introduce an online automatic augmenter (OAA) to diversify data through various image transformations to enhance generalization ability. We propose a residual robustness enhancement (RRE) block to enrich feature diversity and inject perturbations to produce varied inputs for different decoders, therefore enhancing the variability. Furthermore, we introduce a sensitivity loss to improve consistency across varying decoders and stabilize training.
    \item Extensive results from both our own and three public datasets underscore the superiority of DEMS. Notably, DEMS shows increasing performance advantages in extreme data shortages, emphasizing its extraordinary efficiency in severe data shortages.
\end{itemize}

The rest of this paper is organized as follows. Section \hyperref[2]{2} meticulously reviews the related works, focusing on DA and SSL in medical image segmentation. Existing challenges and potential improvements are then summarized. In Section \hyperref[3]{3}, we provide a detailed illustration of the proposed approach, including the introduction of DEMS, OAA, RRE block and connection structure, and loss function. Section \hyperref[4]{4} presents the datasets, experimental setup, and evaluation metrics. In Section \hyperref[5]{5}, we illustrate the experimental results, conduct comprehensive analysis and visualization, and perform intensive ablation experiments. We conclude our work and provide insightful future perspectives in Section \hyperref[6]{6}.


\section{Related Work}
\label{2}

\subsection{Data Augmentation}
\label{2.1}

Numerous researchers leverage conventional DA as a component in their proposed methods. For instance, Li et al. \citep{li2023hal} presented a HAL-IA method to reduce the annotation cost, in which Gaussian noise, random flip, and random crop were leveraged for DA. Yu and colleagues \citep{yu2023unest} developed a UNesT network for local communication, in which various DA transformations such as random flip, rotation, and intensity change were leveraged to train the renal segmentation model. Li et al. \citep{li2023global} introduced a GFUNet that integrates Fourier transform with U-Net and leverages varying transformations including random histogram matching, rotation, and shifting to perform DA. Zhao and colleagues \citep{zhao2023mskd} developed a novel knowledge distillation framework, in which DA transformations including random scaling, random rotation, and random elastic deformation were utilized. Furthermore, Isensee et al. \citep{isensee2021nnu} presented a nnU-Net with a preset DA pipeline that sequentially applies transformations such as rotation, scaling, Gaussian noise, Gaussian blur, etc. The conventional DA is intuitive and effective while the design of the augmentation pipeline predominantly relies on experience.

Leveraging GAN to synthesize artificial images and enlarge datasets for DA presents an effective alternative. It generates images at the pixel level, thus being considered more versatile compared with the conventional DA. In 2022, Chai et al. \citep{chai2022synthetic} developed a DPGAN consisting of three variational auto-encoder GANs to synthesize artificial images and labels. Moreover, an extra discriminator was leveraged to promote the image reality and relationship across images and latent vectors. Li and colleagues \citep{li2021semantic} proposed a semi-supervised framework to capture the joint image-label distribution and synthesize chest X-ray and liver computed tomography images. Kugelman et al. \citep{kugelman2020dual} leveraged Style-GAN \citep{karras2019style} to create image-mask pairs for optical coherence tomography images, in which the image and mask are generated on adjacent channels. Pandey and colleagues \citep{pandey2020image} developed a two-stage GAN approach, in which the mask is initially synthesized, followed by the image leveraging the generated mask. Besides, Iqbal et al. \citep{iqbal2023unet} presented a data expansion network in 2023 to synthesize images solely and leveraged a trained convolutional neural network (CNN) to label the generated images. The GAN-based DA faces varied challenges such as varying synthesizing quality, extensive training consumption, and high data appetite.

The automatic DA has garnered significant attention for its extraordinary augmentation diversity. Various approaches adapt reinforcement learning to perform automatic DA. For example, Yang et al. \citep{yang2019searching} leveraged the validation accuracy to update the recurrent neural network controller. The AADG framework presented by Lyu and colleagues \citep{lyu2022aadg} introduces a novel proxy task, in which Sinkhorn distance was utilized to maximize the diversity across various augmented domains. Qin et al. \citep{qin2020automatic} proposed a joint-learning strategy that combines Dueling DQN \citep{wang2016dueling} and segmentation modules to search for maximum performance improvement. Xu and colleagues \citep{xu2020automatic} devised a differentiable approach to update the parameters using stochastic relaxation and the Monte Carlo method. Besides the methods based on reinforcement learning, the MedAugment \citep{liu2023medaugment} developed in 2023 augments each input image with different transformations sampled from two groups of transformations. Zhao et al. \citep{zhao2019data} presented a novel automatic DA method, in which spatial transform and appearance transform were leveraged to synthesize labeled examples. Additionally, Eaton and colleagues \citep{eaton2018improving} utilized the online mix-up \citep{zhang2018mixup} algorithm to enhance the performance of brain glioma segmentation. These methods arise with various challenges including high computation costs, considerable stage complexity, and potential model confounding.

\subsection{Semi-Supervised Segmentation}
\label{2.2}

The MT-based approaches have been widely adopted in the field of semi-supervised medical image segmentation. Zhang et al. \citep{zhang2023uncertainty} proposed an uncertainty-guided mutual consistency learning framework with two branches for pixel-wise classification and level set function regression, respectively. Lei and colleagues \citep{lei2022semi} presented an ASE-Net comprising both discriminator and segmentation networks. The adversarial consistency training approach leverages two discriminators to extract prior relationships, and the dynamic convolution-based bidirectional attention component is utilized to adaptively adjust the network weights. Xu et al. \citep{xu2021shadow} introduced a shadow-consistent SSL approach featuring shadow augmentation and shadow dropout mechanisms. The shadow augmentation mechanism enhances the samples by integrating simulated shadow artifacts, and the shadow dropout mechanism compels the network to identify the boundary using shadow-free pixels. Lyu and colleagues \citep{lyu2023adaptive} developed an AFAM-Net, which utilizes a reconstruction task to capture anatomical information and an adaptive feature aggregation strategy to transfer and filter features. The unsupervised consistency loss is formulated based on the reconstruction and segmentation tasks. Additionally, Yu et al. \citep{yu2019uncertainty} presented an uncertainty-aware framework in which the teacher model generates target outputs together with the uncertainty of each prediction through Monte Carlo sampling. Although the MT-based methods have demonstrated promising performance in various scenarios, various challenges exist. Firstly, the performance of the teacher network can be constrained. Secondly, the shared structure restricts the model variability.

In addition to the MT-based methods, Tang et al. \citep{tang2023multi} developed an MGCC model that adopts the encoder-decoder architecture and incorporates the multi-scale attention gate for feature enhancement. Wu and colleagues \citep{wu2021semi} proposed an MC-Net consisting of an encoder and two decoders and calculated the prediction discrepancy between decoder outputs to promote mutual consistency. Subsequently, the MC-Net+ with further intra-model diversity was presented \citep{wu2022mutual}, in which an additional decoder with the nearest interpolating operation was incorporated. The ASS-GAN presented by Zhai et al. \citep{zhai2022ass} leverages two generators and a discriminator to perform adversarial learning, in which the predicted masks of a generator were utilized to guide the other. Additionally, Wang and colleagues \citep{wang2023mcf} developed a mutual correction framework, in which a contrastive difference review module was leveraged to replace the potential moving average. Moreover, they introduced a novel rectification loss based on the ratio of potential mispredicted area loss to area size. Luo et al. \citep{luo2022semi} introduced a CTCT that employs both U-Net and Swin-UNet \citep{cao2022swin} to perform cross-teaching between the CNN and transformer. These approaches can face restricted variability and limited training stability.

\subsection{Challenges and Improvements}
\label{2.3}

The primary challenges in achieving high-accuracy medical image segmentation lie in leveraging limited and unlabeled data more effectively through DA and SSL. To leverage limited data, current methods encompass conventional DA, GAN-based DA, and automatic DA. However, each of these approaches faces unique challenges. The design of the conventional DA pipeline heavily relies on empirical knowledge. The GAN-based DA may experience inconsistent synthesis quality, high computational costs, and significant data requirements. The automatic DA can lead to considerable computational overhead, procedural complexity, and potential model confusion. As an alternative, we propose a training-free OAA approach that involves minimal computational costs and eliminates stage complexity. This approach augments input data with tailored transformations during training in an online augmentation manner to enhance generalizability. In terms of utilizing unlabeled data, existing architectures are primarily categorized into MT-based and miscellaneous methods. However, methods based on various architectures encounter varying challenges. MT-based approaches are limited by constrained teacher network performance and model variability. Miscellaneous methods can face limited variability and reduced stability. To this end, we propose a meticulously designed encoder-decoder architecture, incorporating the innovative RRE blocks and sensitivity loss. The RRE block diversifies features and introduces perturbations to yield varied decoder inputs, therefore enhancing variability. The sensitivity loss enhances the consistency across varying decoders and stabilizes the training process. We combine a multi-term loss function with a warming-up function to enhance training stability, especially during the early training stages.


\section{Methods}
\label{3}

\subsection{Data-Efficient Medical Segmenter}
\label{3.1}

\begin{figure*}
	\centering
	  \includegraphics[width=\textwidth]{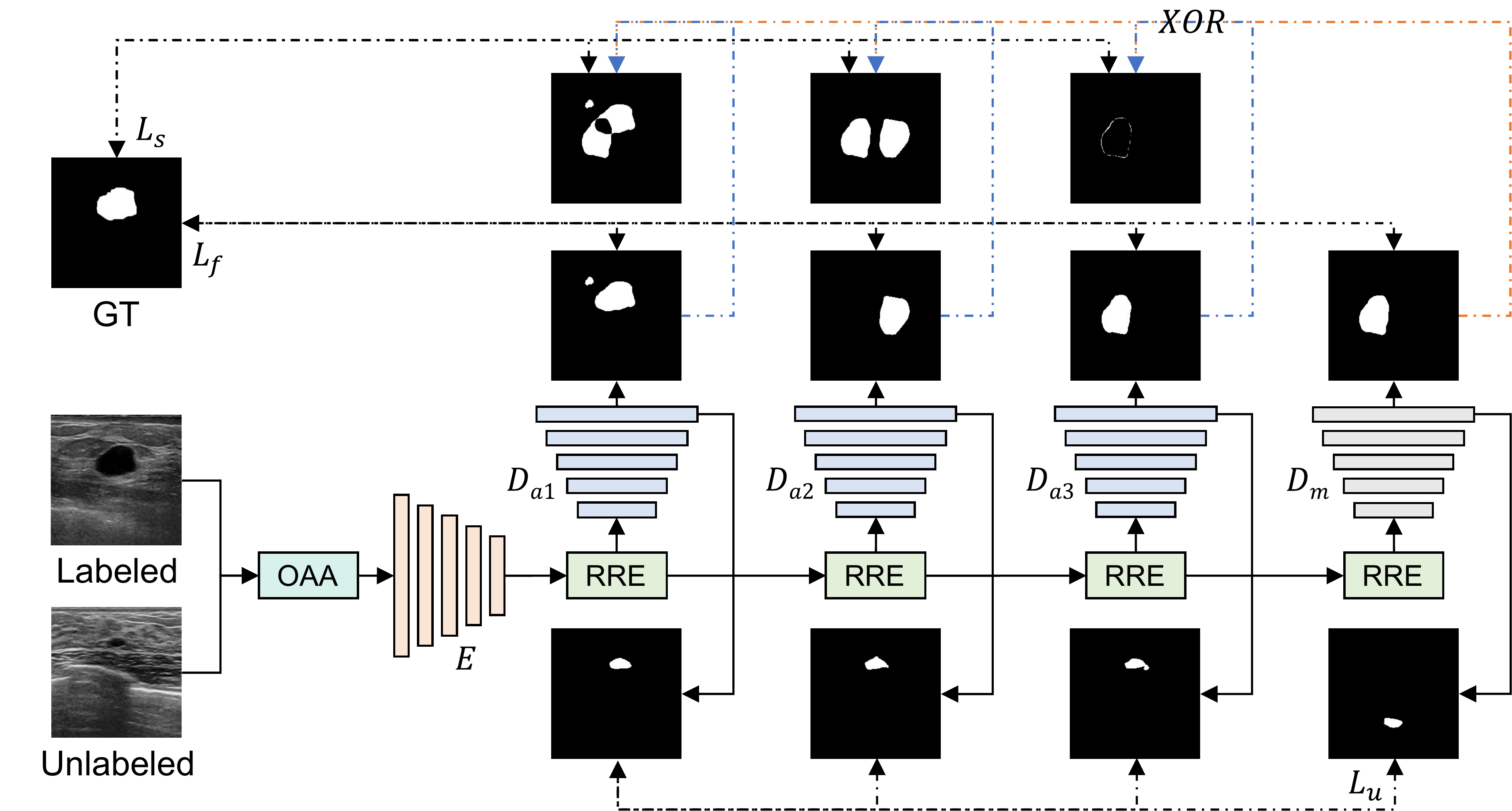}
	\caption{Detailed architecture of DEMS. The DEMS consists of an encoder $E$, a main decoder $D_{m}$, and three auxiliary decoders $D_{a1}$, $D_{a2}$, and $D_{a3}$. It comprises proposed online automatic augmenter (OAA) and residual robustness enhancement (RRE) blocks. The OAA augments the visual input with varying image transformations to diversify the data to enhance generalization ability. The RRE block enriches the feature diversity and introduces perturbations to create varied decoder inputs, thereby bolstering the variability. The loss function comprises fusion loss $L_{f}$, sensitivity loss $L_{s}$, and unsupervised loss $L_{u}$. The introduced sensitivity loss further enhances the consistency across various decoders and stabilizes model training. The $L_{s}$ is formulated based on the XOR operation across the main and auxiliary decoder pairs. The XOR operation is depicted with an overlay of blue and orange dash-dot lines. GT stands for the ground truth.}
	\label{fig1}
\end{figure*}

The architecture of DEMS is depicted in \cref{fig1}. The DEMS utilizes an encoder-decoder architecture and incorporates the developed OAA and RRE blocks. It should be noted that the auxiliary decoders are used exclusively during the training phase. The structure of the encoder $E$, main decoder $D_{m}$, and auxiliary decoders $D_{a1}$, $D_{a2}$, and $D_{a3}$ follows that of the U-Net \citep{ronneberger2015u}. The OAA augments input data with a variety of image transformations, thereby diversifying the dataset to improve the generalization capability. The RRE block diversifies features and introduces perturbations to generate varying decoder inputs, thereby providing enhanced variability. The loss function $L$ consists of fusion loss $L_{f}$ and sensitivity loss $L_{s}$ for labeled images, and unsupervised loss $L_{u}$ for unlabeled images. We propose the sensitive loss to further enhance the consistency across various decoders and stabilize model training.

\subsection{Online Automatic Augmenter}
\label{3.2}

We revisit the fundamental aspects of MedAugment to establish the basis for our proposed improvements. MedAugment is an offline automatic DA approach that expands the input dataset into a larger dataset during a separate preparatory stage ahead of model training. It comprises $N=4$ augmentation branches along with an additional branch. Each augmentation branch performs $M = {[2, 3]}$ transformations on each input. The transformations are sampled from the pixel augmentation space $A_{p}$ and spatial augmentation space $A_{s}$ using one of the sub-strategies in the sampling strategy $\Pi$. The strategy comprises four sub-strategies, sampling $[1, 0, 1, 0]$ and $[2, 3, 1, 2]$ transformations from $A_{p}$ and $A_{s}$, respectively. The sampled transformations for each branch are subsequently shuffled. The maximum magnitude of each transformation $M_{A}$ and the augment probability $P_{A}$ are controlled based on the augmentation level $L$, and the applied magnitude is uniformly sampled within the maximum boundary. The additional branch is maintained to preserve the source visual information.

\begin{algorithm}[!pt] 
\begin{algorithmic}[1]
\caption{Pseudocode of OAA.}
\label{algorithm1}  
\Require Pixel augmentation space $A_{p}$ = \texttt{[brightness, contrast, posterize, sharpness, Gaussian blur, Gaussian noise]}, spatial augmentation space $A_{s}$ = \texttt{[rotate, horizontal flip, vertical flip, scale, x-axis translate, y-axis translate, x-axis shear, y-axis shear]}, number of transformation $M = {[2, 3]}$, sampling strategy $\Pi = {[\pi_1, \pi_2, \pi_3, \pi_4]}$, augmentation level $L=5$, maximum transformation magnitude $M_{A}$, transformation probability $P_{A} = 0.2L$, input image and mask pair $P_{IM}$;
\Ensure Output pair $P_{IM}^{\prime}$;
\State Sample $\pi$ from $\Pi$ 
\State Sample $M$ transformations $O = {[o_1, ..., o_M]}$ using $\pi$ from $A_{p}$, $A_{s}$
\State Shuffle $O$ 
    \ForAll {$o$}
        \State Calculate $M_{A}$, $P_{A}$ using $L$ 
        \State Uniformly sample magnitude $m_{A}$ $\in$ $M_{A}$
    \EndFor
    \State $P_{IM}^{\prime} = OP_{IM}$
\State Out $P_{IM}^{\prime}$
\end{algorithmic}
\end{algorithm}

\begin{figure*}
	\centering
	  \includegraphics[width=0.8\textwidth]{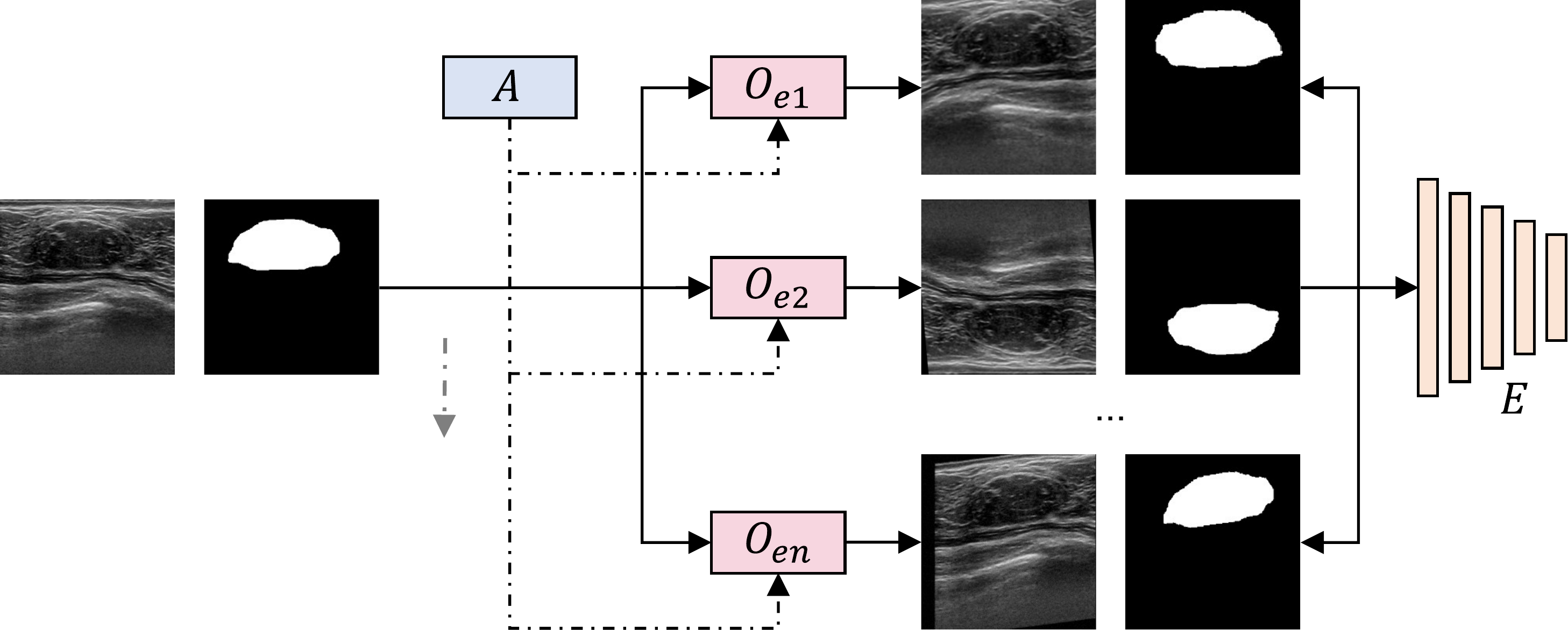}
	\caption{Workflow of the proposed OAA. Each input image and its corresponding mask undergo diverse DA transformations $O$ sampled from augmentation spaces $A$ at each new epoch $e$. The encoder receives diverse inputs in successive epochs as training proceeds, thereby enhancing the generalization capability. The gray dash-dot line depicts the training progress.}
	\label{fig2}
\end{figure*}

We introduce the OAA based on MedAugment to perform online automatic DA and diversify the input data to enhance the generalization capability. We consolidate multiple branches into a single main branch to perform one-to-one augmentation. Following this setup, each input data undergoes augmentation by the transformations sampled using one of the sub-strategies from $\Pi$. We uniformly distribute the probability of employing each sub-strategy. The additional branch is excluded as it does not contribute to the one-to-one augmentation setup. Given the limited number of transformations sampled from $A_{p}$, and considering that transformations in $A_{s}$ do not compromise the effectiveness of medical images, we adopt sampling with replacement to offer enriched data diversity to enhance generalization ability. This modification offers a more comprehensive set of transformation combinations, making the combinations with identical transformations available for sampling. We illustrate the pseudocode of OAA in \Cref{algorithm1} and it demonstrates the process of augmenting an input image and its corresponding mask within a single epoch. It should be noted that several of the transformations do not hold magnitude. In \cref{fig2}, we demonstrate the workflow of OAA, illustrating how an example input data is augmented throughout the training process. The input image and mask undergo diverse transformations at each successive epoch, thereby ensuring remarkable generalization ability. We emphasize that the OAA can be seamlessly integrated into the established DA pipeline \citep{buslaev2020albumentations} with one line of code, and utilizing it is as straightforward as leveraging arbitrary conventional DA transformations.

\subsection{Block and Connection Structure}
\label{3.3}

\begin{figure*}
	\centering
	  \includegraphics[width=\textwidth]{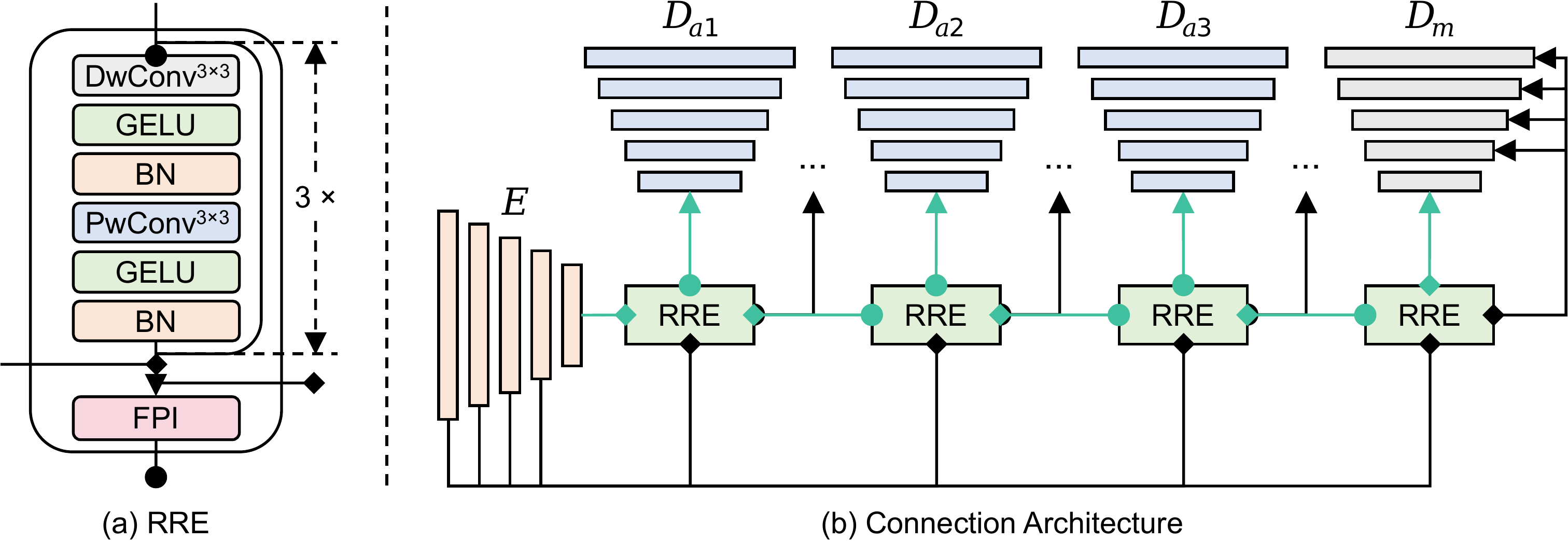}
	\caption{Detailed structure of the RRE block and the connection structure between the encoder and varying decoders. The RRE block features two distinct input-output pairs denoted with rhombus and circle by the shapes of the starting and ending arrows. It mainly encompasses residual connection, depthwise convolution (DwConv), pointwise convolution (PwConv), and feature perturbation injection (FPI) block. We denote the output of the encoder at varying blocks with $f_{1}$, $f_{2}$, $f_{3}$, $f_{4}$, and $f_{5}$. We represent the streams starting at $f_{1}$, $f_{2}$, $f_{3}$, and $f_{4}$ with black arrows, and streams starting at $f_{5}$ with teal arrows. For clarity and conciseness, the skip connections across the encoder and main decoder are exclusively depicted. BN denotes the batch normalization layer.}
	\label{fig3}
\end{figure*}

We develop the RRE block to diversify features and introduce perturbations to produce varied decoder inputs to enhance variability. We illustrate the structure of the RRE block in \cref{fig3}a. The RRE block features two distinct input-output pairs symbolized by rhombus and circle, respectively. The primary components of the RRE block include residual connection \citep{he2016deep}, depthwise convolution (DwConv), pointwise convolution (PwConv), and feature perturbation injection (FPI) block. The residual connection helps train the deep network by mitigating the vanishing gradient problem, thus ensuring a smoother flow of information. The configuration of DwConv and PwConv endows the model with powerful representational capabilities while maintaining computational efficiency. The FPI block injects diverse perturbations encompassed from feature noise \citep{ouali2020semi}, feature dropout \citep{ouali2020semi}, and dropout \citep{tompson2015efficient}.

The connection structure between the encoder and decoders is demonstrated in \cref{fig3}b. The encoder comprises five convolution blocks, with four max pooling layers inserted succeeding each of the first four blocks. Each convolution block consists of two 3 $\times$ 3 convolution layers, followed by a batch normalization layer and the GELU activation function. The decoder is structured into four upsampling stages, each consisting of an upsampling block, a feature concatenation layer, and a convolution block. The upsampling block comprises an upsampling layer, followed by a 3 $\times$ 3 convolution layer, a batch normalization layer, and the GELU activation function. Given the encoder output at varying blocks $f_{1}, f_{2}, f_{3}$, $f_{4}$, and $f_{5}$, the first RRE block receives inputs from all blocks through two distinct input ports. Subsequent blocks take both $f_{1}, f_{2}, f_{3}$, $f_{4}$, and outputs from preceding blocks as inputs. Additionally, the output ports associated with the skip connections between the main decoder and auxiliary decoders differ. The meticulous connection design ensures that each decoder input undergoes a distinct number of convolution operations, diverse perturbations, or both.

\subsection{Loss Function}
\label{3.4}

The fusion loss $L_{f}$ fuses binary cross entropy loss $L_{BCE}$ and dice loss $L_{DSC}$ to ensure robust segmentation performance across various scales. Primarily, the $L_{BCE}$ focuses on pixel-level accuracy, penalizing deviations of predicted pixel values from the actual labels. In contrast, the $L_{DSC}$ emphasizes object-level accuracy by measuring the overlap between the prediction and the ground truth (GT). Given the GT $y$, decoder prediction $\hat{y}$, and pixel index $i$, the $L_{f}$ for each decoder can be calculated using \cref{eq1}:
\begin{align}
L_{f} & = 0.5 \cdot L_{BCE}(\hat{y}, y) + L_{DSC}(\hat{y}, y) \label{eq1} \\
&= -\frac{1}{2N} \sum_{i=1}^N \left[ y_i \cdot \log(\hat{y}_i) + (1-y_i) \cdot \log(1-\hat{y}_i) \right] + 1 - \frac{2 \cdot \sum_{i=1}^{N} \hat{y}_i y_i}{\sum_{i=1}^{N} \hat{y}_i + \sum_{i=1}^{N} y_i} \notag
\end{align}

We present a novel sensitivity loss $L_{s}$ to further bolster the consistency across various decoders and enhance training stability. The formulation of $L_{s}$ is based on an intuitive principle in binary segmentation where a mismatch in predictions across two different decoders suggests an error in one of them. To this end, we term the area predicted by different decoders with various binarized results as the sensitivity area and leverage the sensitive loss to encourage the decoder pairs to reduce the area size. Given the output of the main and arbitrary auxiliary decoder $\hat{y}_{m}$, and $\hat{y}_{a}$, and the binarization threshold $T$ of 0.5, the $L_{s}$ for each $\hat{y}_{m}-\hat{y}_{a}$ pair can be formulated based on XOR operation using \cref{eq2}:
\begin{align}
L_{s} = \frac{1}{N} \sum_{i=1}^{N} (\hat{y}_{m,i} > T) \oplus (\hat{y}_{a,i} > T)
\label{eq2} 
\end{align}

The unsupervised loss $L_{u}$ is meticulously designed to harness the information embedded within unlabeled images. The implementation of $L_{u}$ bears resemblance to that of the sensitivity loss, in which the essence is to boost the consistency across different decoders. We utilize the mean squared error loss to compute the $L_{u}$, and the computation for each $\hat{y}_{m}-\hat{y}_{a}$ pair is formulated as shown in \cref{eq3}:
\begin{align}
L_{u} & = \sum_{i=1}^N (\hat{y}_{m,i}-\hat{y}_{a,i})^{2}
\label{eq3}
\end{align}

The loss functions delineated in \cref{eq1}, \cref{eq2}, and \cref{eq3} compute the loss for each decoder or decoder pair, and the complete loss terms can be formulated based on their mean. Specifically, the \(\bar{L}_{f}\) is derived from the average losses for four individual decoders, while \(\bar{L}_{s}\) and \(\bar{L}_{u}\) are calculated based on the mean losses of three distinct decoder pairs. To reduce the risk of sensitivity and unsupervised losses destabilizing the training, especially in the initial stages, we progressively increase the weight of these two terms as training progresses. Specifically, we employ the Gaussian warming-up function $\lambda$ \citep{yu2019uncertainty} to adjust the weight based on the training step $t$ and maximum training step $t_{max}$ using \cref{eq4}:
\begin{equation}
\lambda(t, t_{max}) = 
\begin{cases} 
1 & t_{max} = 0 \\
\exp\left(-5 \left(1 - \frac{t}{t_{max}}\right)^2\right) & 0 \leq t \leq t_{max}
\end{cases}
\label{eq4}
\end{equation}

Considering the convergence and stability of the training process, we comprehensively integrate the three loss terms along with the Gaussian warming-up function to formulate the overall loss function $L$. The overall loss function embodies the cumulative attributes of each loss term and can be expressed using \cref{eq5}:
\begin{equation}
L = \bar{L}_{f} + \lambda (\bar{L}_{s} + \bar{L}_{u})
\label{eq5}
\end{equation}


\section{Experiments}
\label{4}

\subsection{Datasets}
\label{4.1}

We utilize both our own and three public datasets for performance evaluation. We choose ultrasound (US) datasets as quintessential examples due to their inherent challenges. Specifically, the US images not only suffer from speckle noise and low quality \citep{wu2022semi} but also face significant data shortages. Conducting experiments under these challenging conditions is crucial for producing more convincing and robust results. The designated datasets include the own stomach nasogastric tube (SNGT), public breast ultrasound (BUS) \cite{yap2020breast, yap2017automated}, breast US images (BUSI) \cite{al2020dataset}, and digital database thyroid image (DDTI) \citep{tnscui2020-seg-rank1st, pedraza2015open}.

The SNGT dataset consists of 221 US images with stomach and feeding tubes collected from two aspects. First, we conducted a retrospective search through the institutional imaging database PACS from 1 January 2020 to 31 December 2022 to retrieve existing stomach images, either with or without the feeding tube. Second, we prospectively recruited patients with in-situ feeding tubes for US scanning following the approval from the ethics review board and the acquisition of patient consent. The US scans were performed using standard commercial US machines from manufacturers including General Electric, Siemens, and Toshiba. A curvilinear probe operating at a frequency of 2-8 MHz was utilized to capture the images. The captured images were then cropped and four objects including the liver, stomach, tube, and pancreas were annotated using the open-source tool Labelme \cite{russell2008labelme}. In this study, we employ the SNGT dataset for tube binary segmentation based on our loss function design.

The BUS dataset is collected from the UDIAT Diagnostic Centre of the Parc Taulí Corporation utilizing the Siemens ACUSON Sequoia C512 system. It is composed of 163 images, including 110 images featuring benign lesions and 53 with cancerous masses. The average image resolution of the BUS dataset is 760 $\times$ 570. The BUSI dataset is collected from 600 female patients aged between 25 and 75 years using the LOGIQ E9 and LOGIQ E9 Agile US systems. It encompasses 780 images, of which 437, 210, and 133 are benign, malignant, and normal images, respectively. The average image resolution of the BUSI dataset is approximately 500 $\times$ 500. In this study, we leverage benign and malignant images for image segmentation. The DDTI dataset comprises 637 B-mode thyroid US images and includes a variety of lesions such as thyroiditis, cystic nodules, adenomas, and thyroid cancers. The US images were curated at the IDIME US Department in Colombia, with patient selection based on the TI-RADS description.

\subsection{Setup and Evaluation}
\label{4.2}

The datasets are divided into training and validation subsets in a ratio of 7:3. For semi-supervised methods, we utilize 20\% and 40\% of labeled images for training. Images and masks are preprocessed to a resolution of 224 $\times$ 224. We employ SGD as the optimizer with a base learning rate of 0.01. The momentum equals 0.9 and the weight decay is set to 0.0001. The learning rate is updated using the cosine annealing schedule \citep{loshchilov2016sgdr}. The implemented loss function is shown in \cref{eq5}, in which the $\lambda$ is updated every 150 iterations. The batch size is set to 8 and the maximum training iterations equals 20000. We perform our experiments using AMD Ryzen 5965WX and NVIDIA RTX 4090. We report the mean and standard deviation of three independent runs with varied seeds in percentage. Model performance is evaluated using five evaluation metrics including dice score (DSC), intersection over union (IoU), sensitivity (SEN), precision (PRE), and pixel accuracy (PA). Metric calculations are conducted using true positives (TP), true negatives (TN), false positives (FP), and false negatives (FN), as detailed from \cref{eq6} to \cref{eq10}:
\begin{equation}
D S C=\frac{2 \times TP}{2 \times TP + FP + FN}
\label{eq6}
\end{equation}
\begin{equation}
I o U=\frac{TP}{TP + FP + FN}
\label{eq7}
\end{equation}
\begin{equation}
S E N=\frac{T P}{T P+F N}
\label{eq8}
\end{equation}
\begin{equation}
P R E=\frac{T P}{T P+F P}
\label{eq9}
\end{equation}
\begin{equation}
P A=\frac{T P+T N}{T P+T N+F P+F N}
\label{eq10}
\end{equation}


\section{Results}
\label{5}

\subsection{Metric Comparison}
\label{5.1}

\begin{table}[ht!]
\centering
\caption{Performance of DEMS and SOTA methods on the SNGT dataset using 20\% and 40\% labeled data. We report the mean and standard deviation of three independent runs in percentage. The best results are highlighted in bold.}
\resizebox{\linewidth}{!}{
\begin{tabular*}{610pt}{ccccccccc}
\toprule
    Method                             & Venue    & Lableded    & Unlabeled  & DSC                     & IoU                     & SEN                     & PRE                     & PA                      \\
\midrule
    U-Net \citep{ronneberger2015u}     & MICCAI   & 30 (20\%)   & 0          & 37.74$\pm$2.50          & 27.78$\pm$1.17          & 44.84$\pm$6.33          & 40.87$\pm$1.83          & 99.09$\pm$0.10          \\
    U-Net                              & MICCAI   & 61 (40\%)   & 0          & 47.47$\pm$0.92          & 36.96$\pm$1.05          & 47.10$\pm$1.20          & 57.46$\pm$2.25          & 99.31$\pm$0.02          \\
\midrule
    CCT \citep{ouali2020semi}          & CVPR     & 30 (20\%)   & 124 (80\%) & 34.27$\pm$1.32          & 25.73$\pm$0.87          & 35.35$\pm$3.89          & 42.08$\pm$1.61          & 99.22$\pm$0.08          \\  
    UA-MT \citep{yu2019uncertainty}    & MICCAI   &             &            & 33.60$\pm$0.61          & 25.65$\pm$0.62          & 34.57$\pm$1.74          & 41.87$\pm$1.56          & 99.25$\pm$0.03          \\  
    MC-Net+ \citep{wu2022mutual}       & MEDIA    &             &            & 38.68$\pm$0.61          & 29.69$\pm$0.57          & 38.69$\pm$0.92          & 46.99$\pm$2.36          & 99.29$\pm$0.02          \\
    MGCC \citep{tang2023multi}         & ArXiv    &             &            & 44.22$\pm$0.01          & 33.62$\pm$0.64          & 50.69$\pm$2.06          & 47.41$\pm$1.96          & 99.18$\pm$0.06          \\
    CPS \citep{chen2021semi}           & CVPR     &             &            & 33.66$\pm$1.62          & 25.50$\pm$0.74          & 34.97$\pm$3.85          & 39.48$\pm$0.56          & 99.23$\pm$0.04          \\  
    CTCT \citep{luo2022semi}           & PMLR     &             &            & 41.83$\pm$1.08          & 31.82$\pm$0.66          & 44.28$\pm$2.37          & 50.64$\pm$1.17          & 99.32$\pm$0.00          \\  
    ICT \citep{verma2022interpolation} & NN       &             &            & 34.13$\pm$0.98          & 26.52$\pm$0.73          & 33.61$\pm$1.32          & 41.28$\pm$0.50          & 99.28$\pm$0.02          \\  
    R-Drop \citep{wu2021r}             & NIPS     &             &            & 32.38$\pm$1.14          & 24.79$\pm$0.92          & 32.71$\pm$1.88          & 39.62$\pm$0.64          & 99.27$\pm$0.01          \\  
    URPC \citep{luo2021efficient}      & MICCAI   &             &            & 35.42$\pm$0.90          & 27.68$\pm$0.72          & 35.19$\pm$2.72          & 42.92$\pm$1.95          & 99.28$\pm$0.02          \\  
    DEMS (Ours)                        & -        &             &            & \textbf{54.59$\pm$0.44} & \textbf{43.39$\pm$0.54} & \textbf{56.84$\pm$0.64} & \textbf{59.60$\pm$2.41} & \textbf{99.32$\pm$0.03} \\
\midrule
    CCT \citep{ouali2020semi}          & CVPR     & 61 (40\%)   & 93 (60\%)  & 42.81$\pm$0.86          & 32.69$\pm$0.69          & 45.92$\pm$1.69          & 49.69$\pm$1.23          & 99.29$\pm$0.02          \\
    UA-MT \citep{yu2019uncertainty}    & MICCAI   &             &            & 46.34$\pm$1.38          & 35.88$\pm$0.77          & 48.38$\pm$2.51          & 54.29$\pm$1.04          & 99.32$\pm$0.02          \\
    MC-Net+ \citep{wu2022mutual}       & MEDIA    &             &            & 48.30$\pm$0.84          & 37.67$\pm$0.71          & 49.67$\pm$0.84          & 57.95$\pm$2.23          & 99.36$\pm$0.02          \\
    MGCC \citep{tang2023multi}         & ArXiv    &             &            & 53.76$\pm$0.80          & 42.74$\pm$0.70          & 58.18$\pm$1.05          & 57.45$\pm$2.23          & 99.34$\pm$0.02          \\
    CPS \citep{chen2021semi}           & CVPR     &             &            & 42.50$\pm$0.99          & 32.80$\pm$0.71          & 43.00$\pm$1.61          & 53.84$\pm$2.68          & 99.33$\pm$0.02          \\
    CTCT \citep{luo2022semi}           & PMLR     &             &            & 46.01$\pm$1.09          & 35.78$\pm$0.68          & 48.11$\pm$1.25          & 54.69$\pm$2.16          & 99.34$\pm$0.01          \\
    ICT \citep{verma2022interpolation} & NN       &             &            & 46.47$\pm$1.69          & 36.20$\pm$1.18          & 48.48$\pm$2.19          & 55.17$\pm$2.15          & 99.34$\pm$0.01          \\
    R-Drop \citep{wu2021r}             & NIPS     &             &            & 38.68$\pm$1.04          & 29.79$\pm$0.75          & 38.10$\pm$0.89          & 49.51$\pm$1.63          & 99.32$\pm$0.01          \\
    URPC \citep{luo2021efficient}      & MICCAI   &             &            & 44.09$\pm$0.45          & 34.58$\pm$0.57          & 45.46$\pm$0.75          & 51.65$\pm$0.74          & 99.35$\pm$0.01          \\
    DEMS (Ours)                        & -        &             &            & \textbf{59.66$\pm$0.21} & \textbf{48.21$\pm$0.62} & \textbf{61.48$\pm$1.01} & \textbf{64.52$\pm$0.46} & \textbf{99.37$\pm$0.02} \\
\bottomrule
\end{tabular*}
}
\label{tab1}
\end{table}

\begin{table}[ht!]
\centering
\caption{Performance of DEMS and SOTA methods on the BUS dataset using 20\% and 40\% labeled data.}
\resizebox{\linewidth}{!}{
\begin{tabular*}{610pt}{ccccccccc}
\toprule
    Method                             & Venue    & Lableded    & Unlabeled & DSC                     & IoU                     & SEN                     & PRE                     & PA                      \\
\midrule
    U-Net \citep{ronneberger2015u}     & MICCAI   & 22 (20\%)   & 0         & 61.68$\pm$1.60          & 51.24$\pm$1.13          & 62.42$\pm$3.26          & 71.60$\pm$3.82          & 96.96$\pm$0.35          \\
    U-Net                              & MICCAI   & 45 (40\%)   & 0         & 74.02$\pm$1.76          & 65.04$\pm$1.34          & 72.54$\pm$3.13          & 84.23$\pm$0.90          & 97.95$\pm$0.03          \\
\midrule
    CCT \citep{ouali2020semi}          & CVPR     & 22 (20\%)   & 92 (80\%) & 61.91$\pm$1.17          & 52.62$\pm$0.81          & 58.64$\pm$0.87          & \textbf{77.89$\pm$1.43} & 96.97$\pm$0.07          \\
    UA-MT \citep{yu2019uncertainty}    & MICCAI   &             &           & 62.09$\pm$1.14          & 52.77$\pm$0.75          & 61.69$\pm$2.38          & 73.63$\pm$2.28          & 97.08$\pm$0.06          \\
    MC-Net+ \citep{wu2022mutual}       & MEDIA    &             &           & 65.92$\pm$0.55          & 56.56$\pm$0.54          & 64.26$\pm$1.78          & 76.62$\pm$1.65          & 97.27$\pm$0.09          \\
    MGCC \citep{tang2023multi}         & ArXiv    &             &           & 72.35$\pm$0.88          & 60.72$\pm$0.83          & 73.05$\pm$3.90          & 78.64$\pm$3.17          & 97.64$\pm$0.03          \\
    CPS \citep{chen2021semi}           & CVPR     &             &           & 55.42$\pm$0.63          & 45.74$\pm$0.57          & 53.39$\pm$0.90          & 70.38$\pm$2.01          & 96.67$\pm$0.06          \\
    CTCT \citep{luo2022semi}           & PMLR     &             &           & 63.86$\pm$1.77          & 54.95$\pm$1.85          & 62.50$\pm$1.47          & 76.67$\pm$2.73          & 97.21$\pm$0.07          \\
    ICT \citep{verma2022interpolation} & NN       &             &           & 63.82$\pm$0.74          & 54.44$\pm$0.99          & 62.91$\pm$1.03          & 76.98$\pm$2.22          & 97.23$\pm$0.08          \\
    R-Drop \citep{wu2021r}             & NIPS     &             &           & 54.90$\pm$1.25          & 46.76$\pm$0.80          & 52.16$\pm$0.50          & 72.54$\pm$5.88          & 96.72$\pm$0.15          \\
    URPC \citep{luo2021efficient}      & MICCAI   &             &           & 59.74$\pm$0.95          & 51.13$\pm$1.02          & 57.38$\pm$0.37          & 72.35$\pm$2.37          & 96.87$\pm$0.09          \\
    DEMS (Ours)                        & -        &             &           & \textbf{76.14$\pm$0.03} & \textbf{65.59$\pm$0.04} & \textbf{81.12$\pm$0.84} & 77.80$\pm$0.85          & \textbf{97.69$\pm$0.01} \\
\midrule
    CCT \citep{ouali2020semi}          & CVPR     & 45 (40\%)   & 69 (60\%) & 76.15$\pm$0.52          & 67.48$\pm$0.58          & 74.74$\pm$0.57          & 85.17$\pm$1.39          & 97.93$\pm$0.08          \\
    UA-MT \citep{yu2019uncertainty}    & MICCAI   &             &           & 73.96$\pm$1.02          & 65.87$\pm$0.81          & 72.11$\pm$1.07          & 84.53$\pm$0.68          & 97.98$\pm$0.13          \\
    MC-Net+ \citep{wu2022mutual}       & MEDIA    &             &           & 76.10$\pm$0.66          & 67.67$\pm$0.62          & 73.50$\pm$0.93          & 86.66$\pm$2.78          & 98.08$\pm$0.07          \\
    MGCC \citep{tang2023multi}         & ArXiv    &             &           & 82.99$\pm$1.05          & 73.98$\pm$0.99          & 81.84$\pm$3.22          & \textbf{88.96$\pm$2.10} & 98.33$\pm$0.08          \\
    CPS \citep{chen2021semi}           & CVPR     &             &           & 69.83$\pm$0.74          & 61.81$\pm$0.68          & 67.05$\pm$1.17          & 83.15$\pm$1.51          & 97.74$\pm$0.06          \\
    CTCT \citep{luo2022semi}           & PMLR     &             &           & 75.11$\pm$0.98          & 66.85$\pm$0.74          & 72.12$\pm$1.24          & 85.50$\pm$1.58          & 98.14$\pm$0.03          \\
    ICT \citep{verma2022interpolation} & NN       &             &           & 74.89$\pm$0.37          & 66.63$\pm$0.66          & 73.06$\pm$0.78          & 86.14$\pm$2.30          & 98.02$\pm$0.06          \\
    R-Drop \citep{wu2021r}             & NIPS     &             &           & 65.59$\pm$0.66          & 57.77$\pm$0.79          & 62.15$\pm$1.18          & 80.17$\pm$3.20          & 97.30$\pm$0.12          \\
    URPC \citep{luo2021efficient}      & MICCAI   &             &           & 71.22$\pm$0.49          & 62.52$\pm$0.57          & 67.11$\pm$0.73          & 83.26$\pm$1.15          & 97.76$\pm$0.01          \\
    DEMS (Ours)                        & -        &             &           & \textbf{83.03$\pm$1.42} & \textbf{74.13$\pm$1.66} & \textbf{85.11$\pm$0.61} & 84.72$\pm$1.21          & \textbf{98.47$\pm$0.05} \\
\bottomrule
\end{tabular*}
}
\label{tab2}
\end{table}

\begin{table}[ht!]
\centering
\caption{Performance of DEMS and SOTA methods on the BUSI dataset using 20\% and 40\% labeled data.}
\resizebox{\linewidth}{!}{
\begin{tabular*}{610pt}{ccccccccc}
\toprule
    Method                             & Venue    & Lableded    & Unlabeled  & DSC                     & IoU                     & SEN                     & PRE                     & PA                      \\
\midrule
    U-Net \citep{ronneberger2015u}     & MICCAI   & 90 (20\%)   & 0          & 70.49$\pm$0.82          & 60.73$\pm$0.70          & 74.32$\pm$1.36          & 75.07$\pm$1.01          & 93.45$\pm$0.03          \\
    U-Net                              & MICCAI   & 180 (40\%)  & 0          & 76.18$\pm$0.61          & 66.56$\pm$0.60          & 78.08$\pm$2.22          & 80.24$\pm$2.10          & 94.81$\pm$0.10          \\
\midrule
    CCT \citep{ouali2020semi}          & CVPR     & 90 (20\%)   & 362 (80\%) & 70.04$\pm$0.67          & 60.82$\pm$0.68          & 70.48$\pm$0.81          & 77.30$\pm$0.45          & 93.98$\pm$0.09          \\
    UA-MT \citep{yu2019uncertainty}    & MICCAI   &             &            & 69.09$\pm$0.43          & 59.68$\pm$0.53          & 70.69$\pm$0.71          & 76.14$\pm$1.36          & 93.72$\pm$0.17          \\
    MC-Net+ \citep{wu2022mutual}       & MEDIA    &             &            & 69.78$\pm$1.05          & 60.71$\pm$0.67          & 69.89$\pm$2.63          & 77.99$\pm$0.65          & 93.94$\pm$0.09          \\
    MGCC \citep{tang2023multi}         & ArXiv    &             &            & 75.10$\pm$0.70          & 65.95$\pm$0.95          & 76.71$\pm$1.38          & \textbf{79.06$\pm$0.72} & 94.40$\pm$0.25          \\
    CPS \citep{chen2021semi}           & CVPR     &             &            & 67.59$\pm$0.73          & 58.77$\pm$0.65          & 68.14$\pm$0.30          & 75.84$\pm$1.44          & 93.68$\pm$0.07          \\
    CTCT \citep{luo2022semi}           & PMLR     &             &            & 69.66$\pm$0.76          & 60.82$\pm$0.62          & 69.94$\pm$1.23          & 77.97$\pm$0.94          & 93.98$\pm$0.07          \\
    ICT \citep{verma2022interpolation} & NN       &             &            & 69.31$\pm$0.19          & 59.69$\pm$0.55          & 69.91$\pm$0.54          & 77.45$\pm$0.94          & 93.84$\pm$0.17          \\
    R-Drop \citep{wu2021r}             & NIPS     &             &            & 64.71$\pm$0.77          & 55.99$\pm$0.81          & 64.91$\pm$0.61          & 75.61$\pm$2.00          & 93.29$\pm$0.09          \\
    URPC \citep{luo2021efficient}      & MICCAI   &             &            & 68.91$\pm$0.97          & 59.92$\pm$0.78          & 68.49$\pm$1.34          & 77.54$\pm$0.64          & 94.02$\pm$0.00          \\
    DEMS (Ours)                        & -        &             &            & \textbf{76.01$\pm$0.54} & \textbf{66.84$\pm$0.55} & \textbf{77.76$\pm$0.26} & 79.51$\pm$1.12          & \textbf{94.47$\pm$0.22} \\
\midrule
    CCT \citep{ouali2020semi}          & CVPR     & 180 (40\%)  & 272 (60\%) & 72.09$\pm$0.63          & 62.79$\pm$0.68          & 72.95$\pm$1.90          & 78.29$\pm$0.83          & 94.38$\pm$0.03          \\
    UA-MT \citep{yu2019uncertainty}    & MICCAI   &             &            & 72.50$\pm$0.65          & 62.91$\pm$0.82          & 74.50$\pm$0.51          & 77.29$\pm$1.51          & 94.32$\pm$0.23          \\
    MC-Net+ \citep{wu2022mutual}       & MEDIA    &             &            & 73.01$\pm$0.71          & 63.88$\pm$0.68          & 75.43$\pm$0.50          & 77.66$\pm$1.11          & 94.55$\pm$0.22          \\
    MGCC \citep{tang2023multi}         & ArXiv    &             &            & \textbf{79.52$\pm$0.66} & \textbf{70.71$\pm$0.65} & 79.89$\pm$0.86          & \textbf{83.47$\pm$0.79} & \textbf{95.20$\pm$0.20} \\
    CPS \citep{chen2021semi}           & CVPR     &             &            & 70.85$\pm$0.57          & 61.69$\pm$0.59          & 70.72$\pm$1.04          & 78.10$\pm$0.38          & 94.18$\pm$0.25          \\
    CTCT \citep{luo2022semi}           & PMLR     &             &            & 72.13$\pm$0.84          & 62.92$\pm$0.82          & 74.01$\pm$1.75          & 77.06$\pm$0.52          & 94.41$\pm$0.14          \\
    ICT \citep{verma2022interpolation} & NN       &             &            & 72.20$\pm$0.53          & 62.78$\pm$0.60          & 74.75$\pm$2.20          & 77.04$\pm$2.29          & 94.15$\pm$0.38          \\
    R-Drop \citep{wu2021r}             & NIPS     &             &            & 72.06$\pm$0.88          & 62.68$\pm$0.68          & 73.38$\pm$2.10          & 77.97$\pm$1.18          & 94.49$\pm$0.14          \\
    URPC \citep{luo2021efficient}      & MICCAI   &             &            & 71.33$\pm$0.53          & 61.82$\pm$0.67          & 74.06$\pm$1.49          & 75.26$\pm$0.25          & 94.11$\pm$0.07          \\
    DEMS (Ours)                        & -        &             &            & 79.17$\pm$0.19          & 70.09$\pm$0.26          & \textbf{81.04$\pm$1.69} & 82.52$\pm$1.70          & 95.06$\pm$0.38          \\
\bottomrule
\end{tabular*}
}
\label{tab3}
\end{table}

\begin{table}[ht!]
\centering
\caption{Performance of DEMS and SOTA methods on the DDTI dataset using 20\% and 40\% labeled data.}
\resizebox{\linewidth}{!}{
\begin{tabular*}{610pt}{ccccccccc}
\toprule
    Method                             & Venue    & Lableded    & Unlabeled  & DSC                     & IoU                     & SEN                     & PRE                     & PA                      \\
\midrule
    U-Net \citep{ronneberger2015u}     & MICCAI   & 89 (20\%)   & 0          & 67.50$\pm$0.78          & 54.96$\pm$0.99          & 76.25$\pm$1.49          & 68.94$\pm$1.27          & 92.79$\pm$0.25          \\
    U-Net                              & MICCAI   & 178 (40\%)  & 0          & 75.17$\pm$0.83          & 63.70$\pm$0.65          & 81.38$\pm$3.24          & 75.35$\pm$1.73          & 94.29$\pm$0.12          \\
\midrule
    CCT \citep{ouali2020semi}          & CVPR     & 89 (20\%)   & 356 (80\%) & 68.62$\pm$0.72          & 56.71$\pm$0.58          & 77.54$\pm$1.58          & 69.35$\pm$1.93          & 92.75$\pm$0.34          \\
    UA-MT \citep{yu2019uncertainty}    & MICCAI   &             &            & 68.72$\pm$0.56          & 56.72$\pm$0.55          & 76.60$\pm$0.16          & 69.50$\pm$1.20          & 92.99$\pm$0.11          \\
    MC-Net+ \citep{wu2022mutual}       & MEDIA    &             &            & 69.31$\pm$0.58          & 57.60$\pm$0.56          & 75.90$\pm$2.00          & 71.64$\pm$1.17          & 93.22$\pm$0.11          \\
    MGCC \citep{tang2023multi}         & ArXiv    &             &            & 67.71$\pm$0.77          & 54.65$\pm$0.58          & 73.99$\pm$2.90          & 71.25$\pm$1.48          & 92.47$\pm$0.19          \\
    CPS \citep{chen2021semi}           & CVPR     &             &            & 67.50$\pm$0.65          & 55.81$\pm$0.65          & 72.81$\pm$1.18          & 71.10$\pm$1.69          & 92.99$\pm$0.27          \\
    CTCT \citep{luo2022semi}           & PMLR     &             &            & 69.66$\pm$0.80          & 57.86$\pm$0.68          & 77.25$\pm$1.35          & 70.90$\pm$0.80          & 93.18$\pm$0.07          \\
    ICT \citep{verma2022interpolation} & NN       &             &            & 69.33$\pm$0.78          & 57.85$\pm$0.73          & 76.01$\pm$0.69          & 70.72$\pm$1.68          & 93.23$\pm$0.24          \\
    R-Drop \citep{wu2021r}             & NIPS     &             &            & 65.93$\pm$0.79          & 53.86$\pm$0.78          & 72.82$\pm$4.61          & 69.79$\pm$4.57          & 92.56$\pm$0.51          \\
    URPC \citep{luo2021efficient}      & MICCAI   &             &            & 68.73$\pm$0.41          & 56.80$\pm$0.68          & 74.85$\pm$1.58          & 71.45$\pm$1.10          & 93.09$\pm$0.15          \\
    DEMS (Ours)                        & -        &             &            & \textbf{73.90$\pm$0.69} & \textbf{61.86$\pm$0.79} & \textbf{80.22$\pm$0.83} & \textbf{74.72$\pm$0.65} & \textbf{93.87$\pm$0.21} \\
\midrule
    CCT \citep{ouali2020semi}          & CVPR     & 178 (40\%)  & 267 (60\%) & 72.07$\pm$0.79          & 60.86$\pm$0.72          & 79.98$\pm$1.94          & 71.99$\pm$0.27          & 93.83$\pm$0.06          \\
    UA-MT \citep{yu2019uncertainty}    & MICCAI   &             &            & 71.77$\pm$0.59          & 60.88$\pm$0.68          & 77.06$\pm$0.52          & 74.12$\pm$1.10          & 94.06$\pm$0.22          \\
    MC-Net+ \citep{wu2022mutual}       & MEDIA    &             &            & 73.67$\pm$0.69          & 62.75$\pm$0.57          & 78.60$\pm$1.05          & 76.18$\pm$0.48          & 94.35$\pm$0.11          \\
    MGCC \citep{tang2023multi}         & ArXiv    &             &            & 76.53$\pm$0.72          & 65.89$\pm$0.86          & 79.73$\pm$0.67          & \textbf{78.99$\pm$0.73} & 94.84$\pm$0.15          \\
    CPS \citep{chen2021semi}           & CVPR     &             &            & 72.02$\pm$0.88          & 60.87$\pm$0.89          & 76.80$\pm$1.67          & 74.70$\pm$0.28          & 94.10$\pm$0.11          \\
    CTCT \citep{luo2022semi}           & PMLR     &             &            & 73.60$\pm$0.58          & 62.64$\pm$0.63          & 79.45$\pm$0.98          & 74.61$\pm$0.54          & 94.41$\pm$0.17          \\
    ICT \citep{verma2022interpolation} & NN       &             &            & 73.09$\pm$0.51          & 61.84$\pm$0.69          & 79.90$\pm$0.53          & 73.82$\pm$0.63          & 93.97$\pm$0.19          \\
    R-Drop \citep{wu2021r}             & NIPS     &             &            & 68.99$\pm$1.09          & 57.95$\pm$0.78          & 72.33$\pm$1.71          & 74.16$\pm$0.80          & 93.89$\pm$0.21          \\
    URPC \citep{luo2021efficient}      & MICCAI   &             &            & 72.72$\pm$0.59          & 61.89$\pm$0.75          & 78.61$\pm$0.67          & 74.16$\pm$2.06          & 94.20$\pm$0.24          \\
    DEMS (Ours)                        & -        &             &            & \textbf{77.78$\pm$0.66} & \textbf{67.22$\pm$0.43} & \textbf{83.48$\pm$0.46} & 78.12$\pm$0.81          & \textbf{95.03$\pm$0.05} \\
\bottomrule
\end{tabular*}
}
\label{tab4}
\end{table}

We show the performance comparison between DEMS and SOTA methods on the SNGT dataset in \cref{tab1}. As shown, the DEMS outperforms SOTA methods to a large extent on the SNGT dataset, followed by the MGCC. The DEMS achieves a DSC of 54.59\% with merely 30 labeled images for training. This result represents an improvement of 16.85\% and 10.37\% over U-Net and MGCC, respectively. When the number of labeled images increases to 61, the DEMS reaches a DSC of 59.66\%, outperforming the U-Net and MGCC with 12.19\% and 5.90\% DSC leadership, respectively. It should be noted that the uniformly high PA across all methods does not necessarily indicate superior performance but may be influenced by the small size of the object being segmented. In the SNGT dataset, the overall size of the tube is significantly smaller than the background, allowing models to achieve high PA even if categorizing all pixels as background. We illustrate the performance of various methods on the BUS, BUSI, and DDTI datasets in \cref{tab2}, \cref{tab3}, and \cref{tab4}, respectively. From the observations, it is evident that the DEMS outperforms the SOTA methods at varying degrees. As shown in \cref{tab2}, the DEMS reaches a DSC of 76.14\% and 83.03\% using 20\% and 40\% labeled images for training, respectively. Conversely, CCT and MGCC outperform DEMS on PRE  by margins of 0.09\% and 4.24\%, respectively. Regarding the metrics illustrated in \cref{tab3}, the DEMS achieves a DSC of 76.01\% using 90 labeled images for training. The second highest DSC is achieved by MGCC with the highest PRE of 79.06\%. When the number of labeled images equals 180, the DEMS realizes the highest SEN of 81.04\% while slightly underperforming MGCC with a DSC lag of 0.35\%. The MGCC also presents a marginally higher IoU, PRE, and PA. For the results indicated in \cref{tab4}, the DEMS reaches the highest DSC of 73.90\% and 77.78\% when trained using 20\% and 40\% labeled images, respectively. Nevertheless, MGCC reaches the highest PRE of 78.99\% when trained using 40\% labeled images, achieving a 0.87\% leadership compared with DEMS.

\subsection{Analysis and Visualization}
\label{5.2}

\begin{figure*}
	\centering
	  \includegraphics[width=0.9\textwidth]{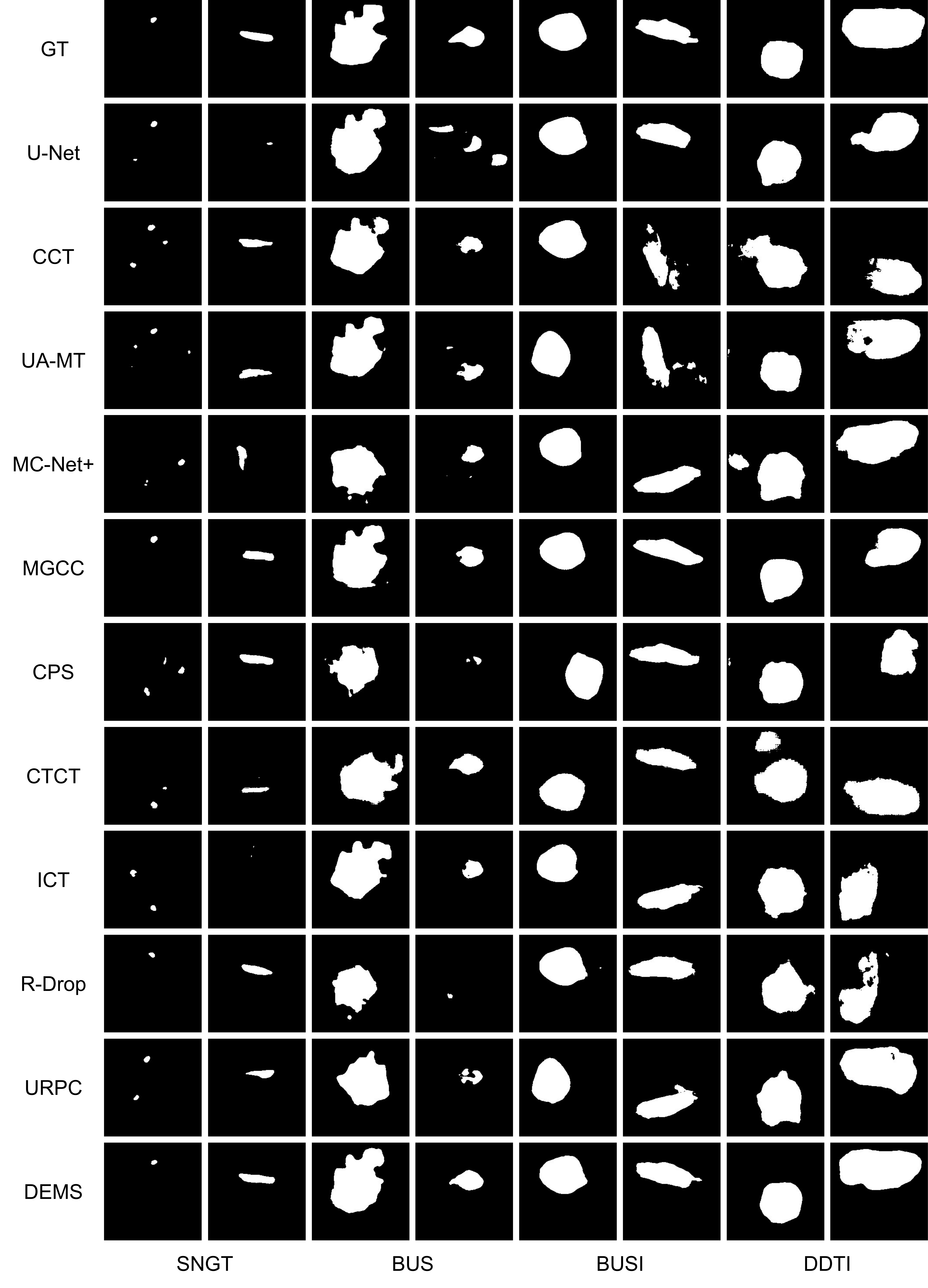}
	\caption{Predicted masks across DEMS and SOTA methods on the four datasets using 40\% labeled data.}
	\label{fig4}
\end{figure*}

\begin{figure*}
	\centering
	  \includegraphics[width=0.9\textwidth]{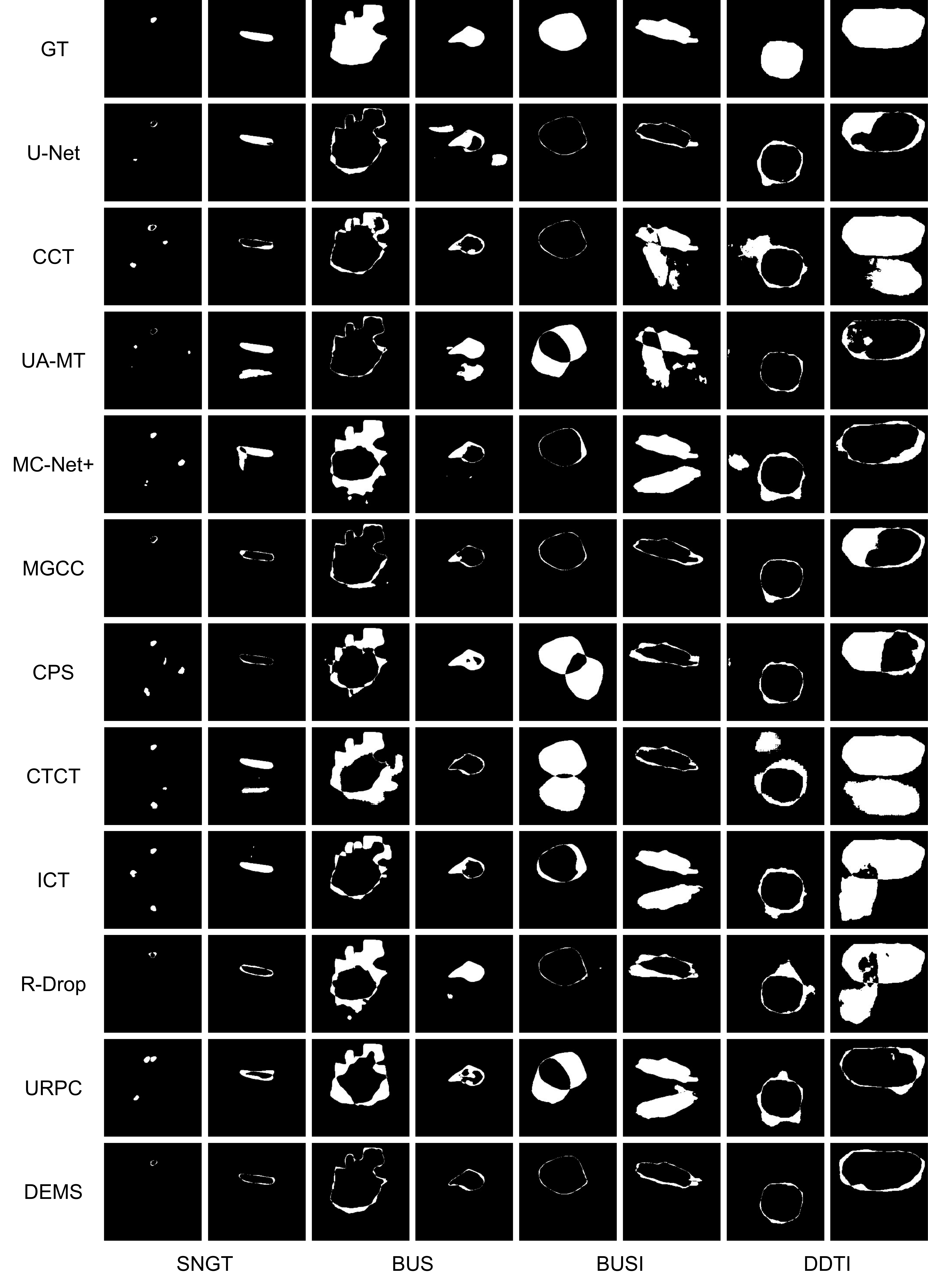}
	\caption{XOR outputs of predicted and GT masks across DEMS and SOTA methods on the four datasets using 40\% labeled data.}
	\label{fig5}
\end{figure*}

We present the predicted masks across DEMS and SOTA methods trained using 40\% labeled images on different datasets in \cref{fig4}. The illustration demonstrates that the masks predicted by DEMS show the most closely resemble compared with the GT. Specifically, the DEMS can predict the contour and the size of the objects most accurately. In the first and second columns that feature small object predictions, most methods such as ICT misidentify non-existent areas and fail to highlight the correct regions. For regular object predictions, although most methods can accurately identify the majority of regions, certain approaches such as CTCT misidentify a large number of areas and can overlook the correct ones. To further illustrate the advantages of DEMS, we demonstrate the XOR outputs between prediction and GT masks across various methods on the four datasets in \cref{fig5}. Observations reveal that DEMS consistently produces the smallest XOR areas in various images, maintaining high prediction accuracy even for objects with complex geometries depicted in the third column. In addition to DEMS, the MGCC also shows desirable performance in most cases except for the examples in the last column. As for the remaining methods, they may accurately predict one or several of the images but often underperform on others. The superior prediction accuracy across varying objects demonstrates the desirable feature capture ability of the developed DEMS.
 
\begin{figure*}
	\centering
	  \includegraphics[width=\textwidth]{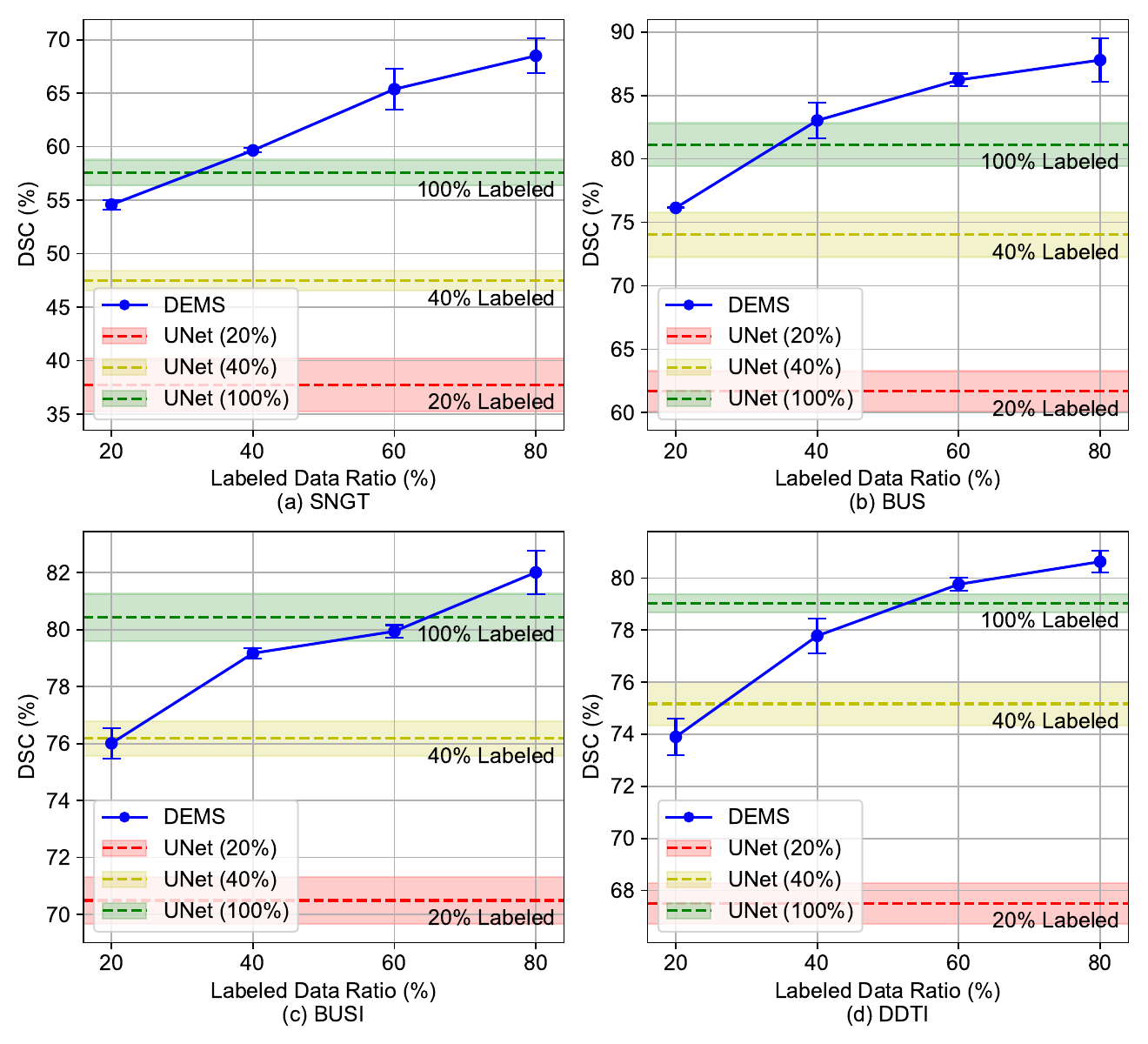}
	\caption{Boundary performance among DEMS and U-Net on the four datasets. The error bar depicts the standard deviation.}
	\label{fig6}
\end{figure*}

To thoroughly investigate the performance of DEMS, we additionally train DEMS using 60\% and 80\% labeled images and compare its performance with the U-Net trained with 100\% labeled images. We term the performance of the U-Net in this setup as the upper bound. Observing \cref{fig6}, it becomes clear that DEMS consistently outperforms the U-Net upper bound across all datasets, achieving the highest DSC of over 68\%, 87\%, 82\%, and 80\% on the SNGT, BUS, BUSI, and DDTI datasets, respectively. Additionally, two patterns are observed. Firstly, DEMS can surpass the upper bound by utilizing a smaller percentage of labeled images on smaller datasets. Specifically, on the smaller SNGT and BUS datasets, the DEMS surpasses the upper bound using merely 40\% labeled images. For the relatively larger BUSI and DDTI datasets, the DEMS exceeds the U-Net upper bound using  80\% and 60\% labeled images for training, respectively. Moreover, the performance advantage is more pronounced on the smaller datasets compared to larger ones. Taking training with 80\% labeled images as an example, the DEMS achieves a DSC superiority of approximately 11\% and 7\% over the U-Net upper bound on the smaller SNGT and BUS datasets. However, this superiority diminishes to about 2\% on the relatively larger BUSI and DDTI datasets. Extensive comparisons across various datasets reveal that DEMS can utilize fewer labeled images to achieve greater performance improvements, showcasing its superior data efficiency.

\begin{figure*}
	\centering
	  \includegraphics[width=\textwidth]{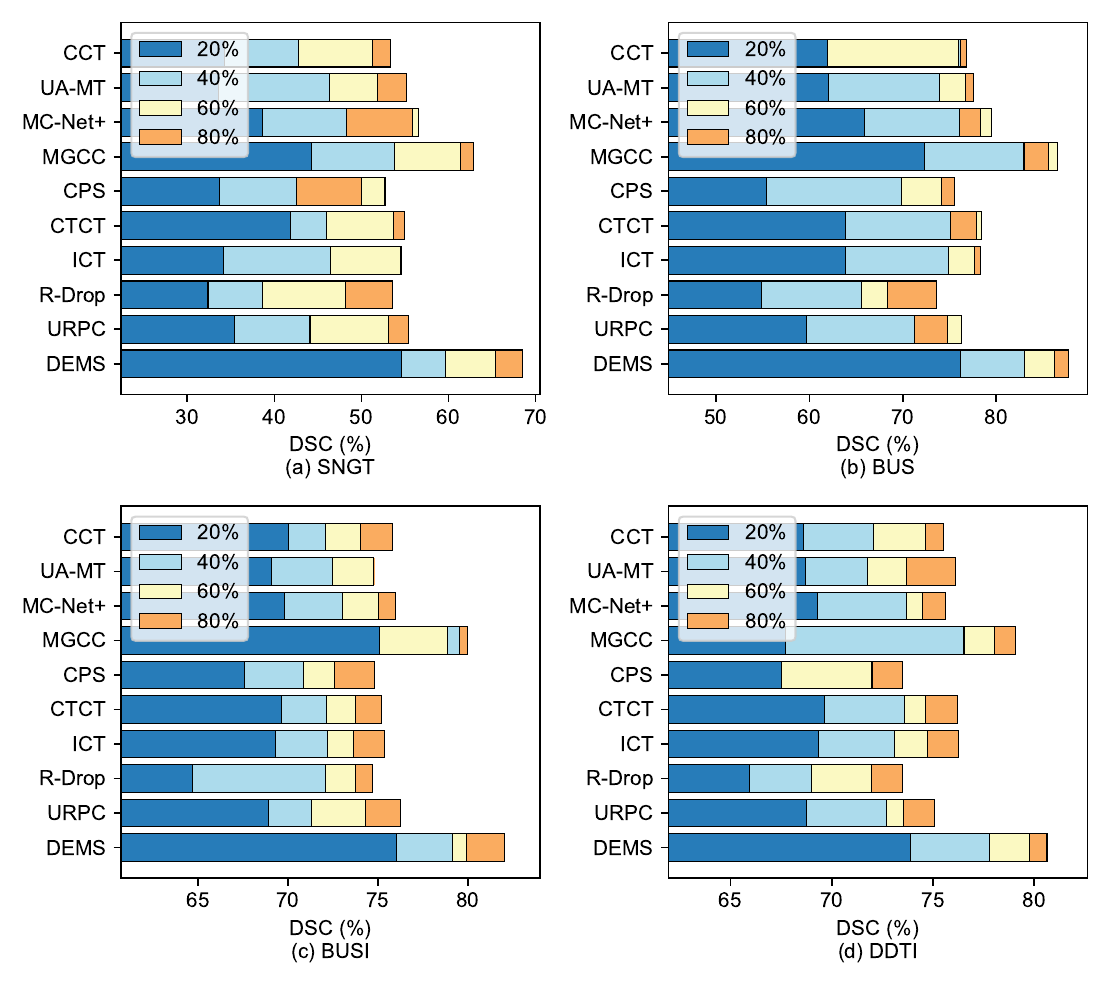}
	\caption{Performance of DEMS and SOTA methods under the complete range of labeled percentages on the four datasets.}
	\label{fig7}
\end{figure*}

Given that MGCC marginally outperforms DEMS, as noted in \cref{tab3}, we conduct extensive supplementary experiments using 60\% and 80\% labeled images to assess the performance across various methods more comprehensively. We show the performance of different methods under the complete range of labeled percentages across four datasets in \cref{fig7}. Through observation, it is evident that the DEMS achieves the highest performance relative to other SOTA methods as the number of labeled images increases. We observe prominent leadership on the SNGT dataset and moderate leadership on the BUS, BUSI, and DDTI datasets. Interestingly, enhanced DSC performance is sometimes observed with fewer labeled images. For instance, the MC-Net+ reaches a higher DSC using 60\% labeled images compared with 80\% labeled images on the SNGT dataset. This phenomenon can be attributed to the limited model stability due to noise susceptibility or overfitting under severe data shortages. This inference aligns with the observations that improved performance with fewer labeled data occurs more frequently on the smaller SNGT and BUS datasets than on the larger BUSI and DDTI datasets. Furthermore, the results depict that performance improvements are more modest on the larger datasets as the percentage of labeled images increases. This outcome is expected, as the model may capture relatively abundant features using a low percentage of labeled images on larger datasets.

\begin{figure*}
	\centering
	  \includegraphics[width=\textwidth]{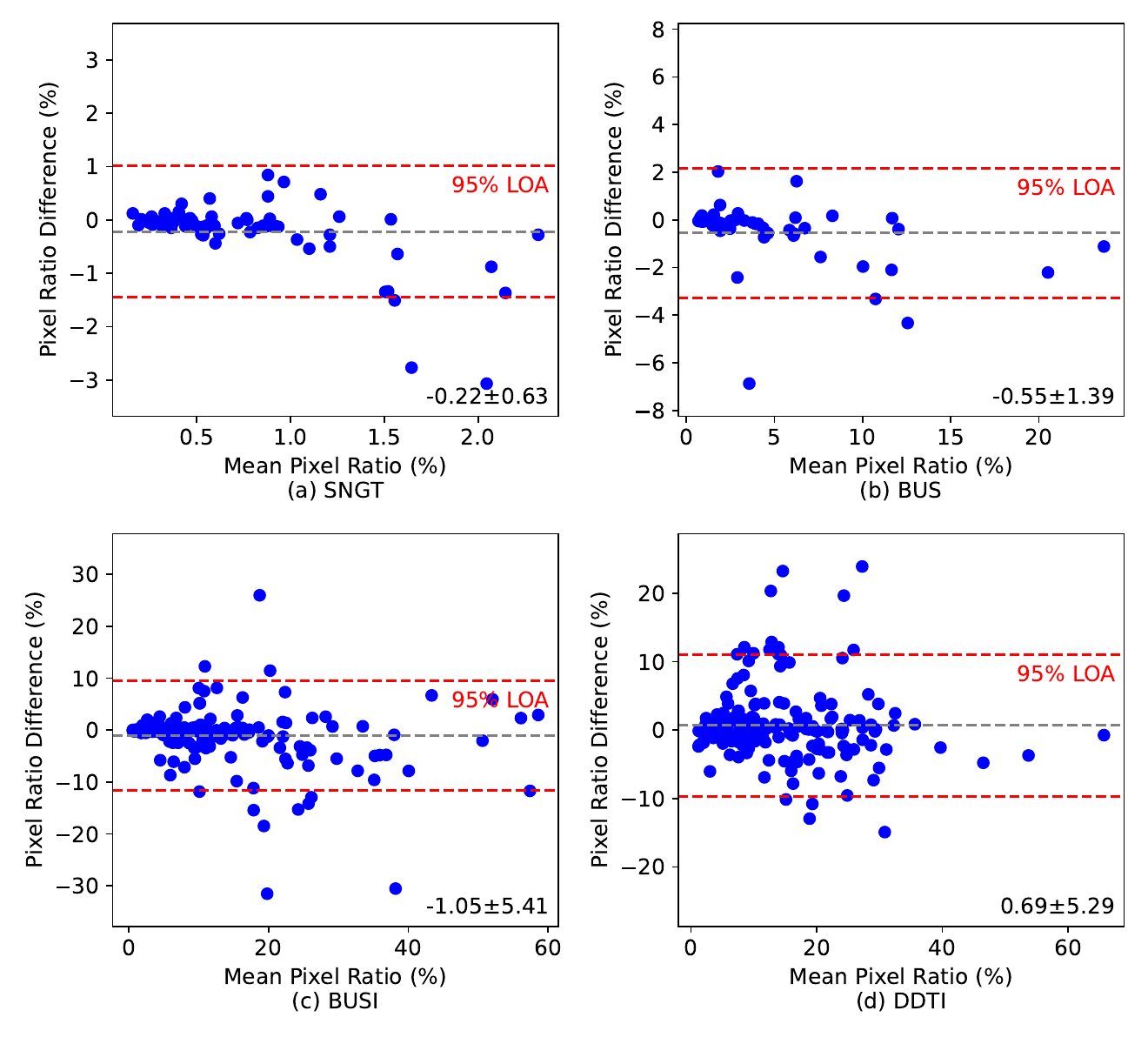}
	\caption{Bland-Altman analysis of DEMS on the four datasets using 80\% labeled images. LOA denotes the 95\% limits of agreement.}
	\label{fig8}
\end{figure*}

We employ the Bland-Altman plot to visualize the prediction consistency for each image in \cref{fig8}. We count the object pixels in the predicted and GT masks, scale each by dividing by $224^2$, and visualize each prediction-GT pair. The analysis clearly shows the predictions of DEMS demonstrate superior consistency in comparison with the GT. The mean differences are notably minimal, hovering close to 0\%, with the maximal and minimal results of 1.05\% and 0.22\% for the BUSI and SNGT datasets, respectively. The standard deviation lies within an ideal range and varies following the actual size of various objects. Specifically, the highest and lowest standard deviations of 5.41\% and 0.63\% are achieved on the BUSI and SNGT datasets, respectively. Furthermore, most points fall within the 95\% limits of agreement, indicating relatively high consistency between the predicted and GT masks. It should be noted that several outliers are observed in each dataset, indicating the presence of inaccuracies in the prediction of these masks. This observation meets our expectations, as datasets containing hundreds of images might include patterns that are evident in the validation subset but absent or underrepresented in the training subset. Additionally, the datasets exhibit no uniform trend in differences, suggesting an absence of a strict systematic pattern in the observed discrepancies. An exception to this is the SNGT dataset, in which the difference increases with the escalation of the mean pixel ratio. This can be attributed to the fact that most of the tubes in the SNGT dataset are substantially small, thereby compelling the model to underpredict the size of the larger tubes. This assumption is consistent with the observation that the outliers predominantly fall along the negative axis. In this scenario, the inconsistency between predictions and GT is predominantly influenced by the dataset-specific attributes rather than the inherent constraints of DEMS.

\begin{figure*}
	\centering
	  \includegraphics[width=\textwidth]{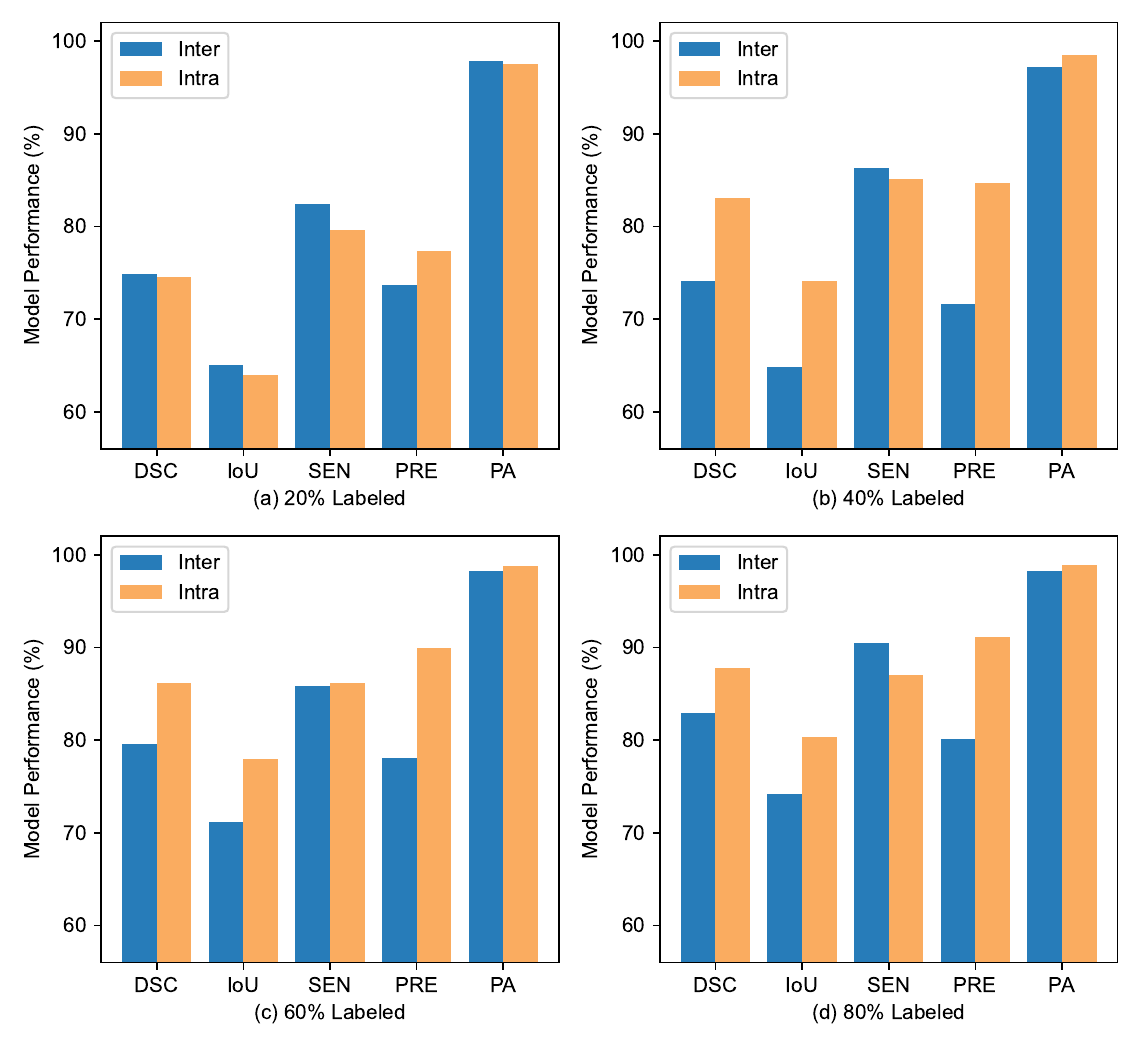}
	\caption{Performance of DEMS on cross-dataset evaluation experiments. The DEMS is trained on the BUS dataset and evaluated on the BUSI and BUS datasets for inter-dataset and intra-dataset configurations, respectively.}
	\label{fig9}
\end{figure*}

We conduct cross-dataset evaluation experiments to assess the generalization capability of the proposed DEMS. The DEMS is trained on the BUS dataset and evaluated on the BUSI dataset for inter-dataset experimentation, with the results being compared to those from intra-dataset experiments. As shown in \cref{fig9}, the DEMS exhibits superior generalization ability and achieves relatively minor performance decreases across various metrics. When trained with 20\% labeled images, DEMS achieves comparable performance in both inter-dataset and intra-dataset settings. This observation may be attributed to the fact that insufficient labeled data does not adequately highlight significant differences between the two datasets. With an increase in the number of labeled images, the performance gap widens and the largest gap is observed when training the DEMS with 40\% labeled data. In this scenario, we observe a performance decrease of around 9\%, 9\%, 13\%, and 1\% in DSC, IoU, PRE, and PA, respectively. Conversely, the SEN shows an approximate 1\% improvement in performance. As the number of labeled training images increases, the performance disparity between inter-dataset and intra-dataset experiments diminishes. This indicates improved generalization capability as additional visual features are incorporated. When leveraging 80\% labeled images for training, the performance gap in metrics reduces to approximately 5\%, 6\%, 9\%, and 1\% for DSC, IoU, PRE, and PA, respectively. Similarly, the SEN exhibits an around 3\% performance increase. The higher SEN in the inter-dataset experiments compared to the intra-dataset experiments may potentially be attributed to less pronounced features in negative examples within the BUSI dataset, leading to an increased likelihood of negatives being recognized as positives. The relatively minor decline in performance underscores the robust generalization capability of the proposed approach.

\begin{figure*}
	\centering
	  \includegraphics[width=\textwidth]{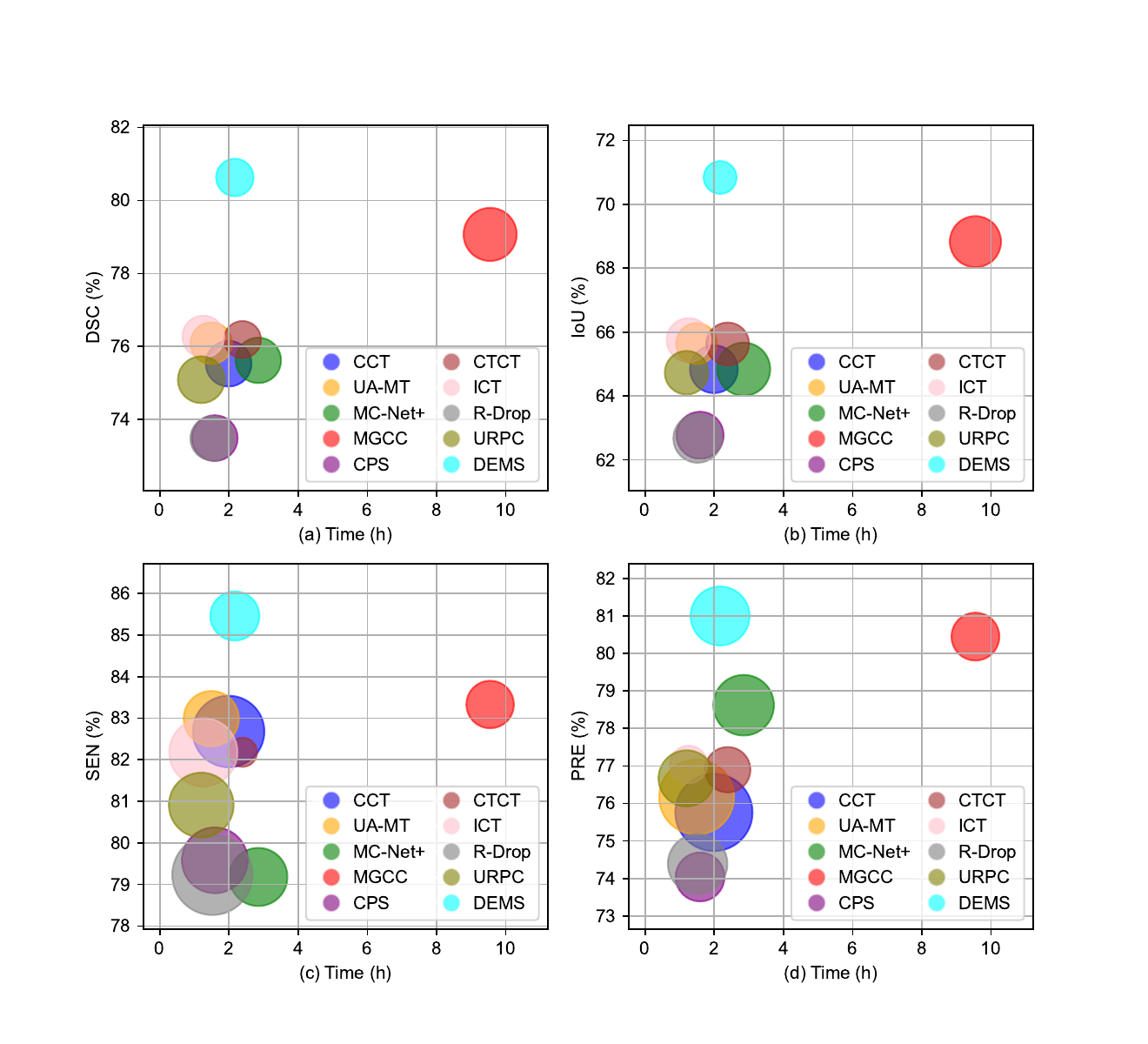}
	\caption{Relationship between time consumption and performance metrics across DEMS and SOTA methods on the DDTI dataset using 80\% labeled images. We exclude PA from the visualization for clarity, as its values are consistently similar across methods and can lead to significant overlap of circles. The circle size indicates the standard deviation.}
	\label{fig10}
\end{figure*}

To evaluate the computational efficiency across models for practical application, we visualize the relationship between model training time and performance metrics across various methods in \cref{fig10}. The results show that DEMS demonstrates superior computational efficiency, achieving the highest evaluation metrics with reasonable time consumption. Most methods require approximately 2 hours to complete training, while MGCC constitutes an exception with a requirement of about 10 hours. DEMS achieves the highest performance, followed by MGCC and other methods. Furthermore, it exhibits relatively low standard deviations, indicating superior stability throughout the training process. Considering both training time and model performance, DEMS not only achieves the highest metrics but also maintains reasonable time consumption. Although MGCC demonstrates relatively high performance, it incurs considerably higher computational costs. The remaining methods consume reasonable training time but fall short of achieving ideal performance. The exceptional computational efficiency of DEMS highlights its potential for practical applications.

\subsection{Ablation Study}
\label{5.3}

\begin{table}[ht!]
\centering
\caption{Ablation study of the OAA, RRE block, and sensitivity loss on the SNGT dataset with 80\% labeled data.}
\resizebox{0.76\linewidth}{!}{
\begin{tabular*}{456pt}{cccccccc}
\toprule
    OAA        & RRE        & $L_{sen}$  & DSC                     & IoU                     & SEN                     & PRE                     & PA                      \\
\midrule
               &            &            & 56.65$\pm$0.71          & 45.32$\pm$0.58          & 58.70$\pm$1.19          & 62.14$\pm$2.07          & 99.32$\pm$0.02          \\
    \checkmark &            &            & 63.50$\pm$1.32          & 52.28$\pm$1.19          & 63.51$\pm$1.52          & \textbf{72.32$\pm$2.32} & 99.43$\pm$0.04          \\
               & \checkmark &            & 62.06$\pm$1.41          & 49.92$\pm$0.93          & 62.93$\pm$2.81          & 70.04$\pm$0.95          & 99.41$\pm$0.02          \\
               &            & \checkmark & 58.42$\pm$0.29          & 46.94$\pm$0.15          & 60.57$\pm$1.81          & 63.58$\pm$2.54          & 99.36$\pm$0.02          \\
    \checkmark & \checkmark &            & 66.83$\pm$1.27          & 55.63$\pm$1.18          & 67.86$\pm$0.86          & 71.60$\pm$3.93          & 99.45$\pm$0.04          \\
    \checkmark &            & \checkmark & 65.12$\pm$0.91          & 54.04$\pm$0.98          & 65.87$\pm$3.13          & 71.82$\pm$3.00          & 99.44$\pm$0.01          \\
               & \checkmark & \checkmark & 63.60$\pm$0.81          & 51.38$\pm$1.09          & 65.57$\pm$1.72          & 68.63$\pm$1.95          & 99.41$\pm$0.04          \\
    \checkmark & \checkmark & \checkmark & \textbf{68.50$\pm$1.64} & \textbf{56.92$\pm$1.22} & \textbf{69.70$\pm$2.32} & 72.30$\pm$3.77          & \textbf{99.46$\pm$0.04} \\
\bottomrule
\end{tabular*}
}
\label{tab5}
\end{table}

To investigate the effectiveness of the OAA, RRE block, and ${L}_{sen}$ within DEMS, we conduct extensive ablation experiments and present the results in \cref{tab5}. It is worth noting that DA transformations including random rotation and random flip are incorporated when OAA is removed. The reason for this is that exclusively removing OAA leads to a drastic deterioration in model performance, thereby rendering comparative analysis futile. Detailed observations reveal that the absence of each component results in a significant performance decrement. Specifically, removing OAA, RRE block, and ${L}_{sen}$ decreases the DSC by 4.90\%, 3.38\%, and 1.67\%, respectively. Additionally, excluding component combinations accelerates performance degradation. When OAA, RRE block, and ${L}_{sen}$ are excluded, a DSC of 56.65\% is observed with a decrease of 11.85\%. Although the variant with solely OAA achieves the highest PRE of 72.32\%, it merely leads by a marginal 0.02\%. However, the DSC, IoU, and SEN significantly fall short by margins of 5.00\%, 4.64\%, and 6.19\%, respectively.


\section{Conclusion}
\label{6}

\subsection{Summary and Discussion}
\label{6.1}

In this manuscript, we introduce a novel semi-supervised segmentation method DEMS to segment medical images with limited data. We devise the OAA to diversify the input data, thereby enhancing the generalization ability. We propose the RRE block to enrich feature diversity and introduce perturbations to produce varying inputs for different decoders, therefore offering greater variability. Furthermore, we propose a novel sensitivity loss to further enhance the consistency across decoders and bolster training stability. Extensive experimental results on both our own and three public datasets demonstrate the superiority of DEMS over SOTA methods. Additionally, DEMS performs exceptionally desirable on severe data shortages, showcasing its remarkable data efficiency and significant advancements in medical segmentation. Despite its impressive performance, adapting DEMS for multi-object segmentation may pose challenges, as its sensitivity loss is formulated specifically for binary segmentation. A potential solution to this limitation involves treating multi-object segmentation as an aggregation of multiple binary segmentation tasks. It should be noted that this kind of solution can come with several challenges such as increased inference times and the complexity of handling overlapping or contiguous objects.

\subsection{Future Perspectives}
\label{6.2}

The future perspectives of the DEMS are twofold. Firstly, an enhanced connection structure between the encoder and decoders can be developed. Recently, transformer-based architecture has been proven to be a powerful tool for computer vision tasks. Compared with CNN-based architecture, it prioritizes the assimilation of global features rather than local features. To this end, incorporating transformer and CNN-based architectures is capable of providing richer visual features and therefore further strengthening the model performance. However, integrating the transformer-based architecture may significantly raise computing costs, potentially limiting the application in most of the mobile scenes. One approach to offset the extra computing lies in reducing the number of decoders. Secondly, a novel updating strategy can be formulated to update the coefficients across various loss function terms. In contrast to fixing the coefficients throughout the training process, adaptively modulating them as training progresses can more effectively maintain the magnitude of each loss term at comparable levels. This can secure the contribution of varying loss terms during the training and is also anticipated to augment the generalization capacity of the model. In addition to the term coefficients, the binarization threshold leveraged to compute the sensitivity loss could also benefit from adaptive adjustments during the training phase.


\section*{Acknowledgements}
This work is supported by Tan Tock Seng Hospital (A-8001334-00-00).



\bibliographystyle{unsrt}
\bibliography{reference.bib}

\biboptions{sort&compress}







\end{sloppypar}
\end{document}